\documentclass[twocolappendix, numberedappendix, iop]{emulateapj}
\usepackage{amsmath,amssymb}
\usepackage{xspace}
\usepackage{graphicx}
\usepackage{graphicx}
\usepackage{epsfig}
\usepackage{hyperref} 
\usepackage{xcolor}
\usepackage{multirow}

\slugcomment{}
\shorttitle{BHAR Dependence on SFR and $M_*$}
\shortauthors{Yang et~al.}

\def\lx{$L_{\rm{X}}$}
\def\lxg{$L_{\rm{X, XRB}}$}
\def\lxgm{$\langle L_{\rm{X, XRB}} \rangle$}

\def\lxm{$\langle L_{\rm X} \rangle$}
\def\bharm{$\langle \mathrm{BHAR} \rangle$}
\def\sfrm{$\langle \mathrm{SFR} \rangle$}

\def\nh{$N_{\rm{H}}$}
\def\med{\mathrm{med}}

\def\chandra{{\it Chandra\/}}

\def\herschel{{\it Herschel\/}}

\def\mbh{$M_{\rm BH}$}

\def\xray{\hbox{X-ray}}  
\def\cdfs{\hbox{CDF-S}}
\def\goodss{\hbox{CANDELS/GOODS-South}}

\newcommand{\varA}[1]{{\operatorname{#1}}}  

\begin{document}

\title{Black-Hole Growth is Mainly Linked to Host-Galaxy Stellar Mass rather than Star Formation Rate}
\author{G. Yang\altaffilmark{1,2}, 
C.-T.~J. Chen\altaffilmark{1,2}, 
F. Vito\altaffilmark{1,2}, 
W.~N. Brandt\altaffilmark{1,2,3}, 
D.~M. Alexander\altaffilmark{4},
B. Luo\altaffilmark{5}, 
M.~Y. Sun\altaffilmark{6}, 
Y.~Q. Xue\altaffilmark{6}, 
F.~E. Bauer\altaffilmark{7,8,9},
A.~M. Koekemoer\altaffilmark{10},
B.~D. Lehmer\altaffilmark{11},
T. Liu\altaffilmark{6},
D.~P. Schneider\altaffilmark{1,2},
O. Shemmer\altaffilmark{12},
J.~R. Trump\altaffilmark{13},
C. Vignali\altaffilmark{14},
J.-X. Wang\altaffilmark{6}
}

\altaffiltext{1}{Department of Astronomy and Astrophysics, 525 Davey Lab, The Pennsylvania State University, University Park, PA 16802, USA; gxy909@psu.edu}
\altaffiltext{2}{Institute for Gravitation and the Cosmos, The Pennsylvania State University, University Park, PA 16802, USA}
\altaffiltext{3}{Department of Physics, 104 Davey Laboratory, The Pennsylvania State University, University Park, PA 16802, USA}
\altaffiltext{4}{Centre for Extragalactic Astronomy, Department of Physics, Durham University, South Road, Durham DH1 3LE, UK}
\altaffiltext{5}{School of Astronomy \& Space Science, Nanjing University, Nanjing 210093, China}
\altaffiltext{6}{CAS Key Laboratory for Research in Galaxies and Cosmology, Center for Astrophysics, Department of Astronomy, University of Science and Technology of China, Chinese Academy of Sciences, Hefei, Anhui 230026, China}
\altaffiltext{7}{Instituto de Astrof{\'{\i}}sica and Centro de Astroingenier{\'{\i}}a, Facultad de F{\'{i}}sica, Pontificia Universidad Cat{\'{o}}lica de Chile, Casilla 306, Santiago 22, Chile}
\altaffiltext{8}{Millennium Institute of Astrophysics (MAS), Nuncio Monse{\~{n}}or S{\'{o}}tero Sanz 100, Providencia, Santiago, Chile}
\altaffiltext{9}{Space Science Institute, 4750 Walnut Street, Suite 205, Boulder, Colorado 80301}
\altaffiltext{10}{Space Telescope Science Institute 3700 San Martin Drive, Baltimore MD 21218, USA}
\altaffiltext{11}{Department of Physics, University of Arkansas, 226 Physics Building, 835 West Dickinson Street, Fayetteville, AR 72701, USA}
\altaffiltext{12}{Department of Physics, University of North Texas, Denton, TX 76203, USA}
\altaffiltext{13}{Department of Physics, 2152 Hillside Road, U-3046, University of Connecticut, Storrs, CT 06269, USA}
\altaffiltext{14}{Universita di Bologn\'{a}, Via Ranzani 1, Bologna, Italy}

\begin{abstract}
We investigate the dependence of 
black-hole accretion rate (BHAR) 
on host-galaxy star formation rate (SFR) 
and stellar mass ($M_*$) in the \goodss\ field in 
the redshift range of $0.5\leq z < 2.0$. 
Our sample consists of $\approx 18000$\ 
galaxies, allowing us to probe galaxies with 
$0.1 \lesssim \mathrm{SFR} \lesssim 100\ M_\sun\ 
\mathrm{yr}^{-1}$ and/or 
$10^8 \lesssim M_* \lesssim 10^{11}\ M_{\sun}$.
We use sample-mean BHAR to approximate 
long-term average BHAR.
Our sample-mean BHARs are derived from 
the \chandra\ Deep Field-South 7~Ms observations,
while the SFRs and $M_*$\ have been estimated 
by the CANDELS team through SED fitting. 
The average BHAR is correlated positively with
both SFR and $M_*$, and the BHAR-SFR and BHAR-$M_*$\
relations can both be described acceptably by 
linear models with a slope of unity.
However, BHAR appears to be correlated more 
strongly with $M_*$\ than SFR. This result indicates that
$M_*$\ is the primary host-galaxy 
property related to black-hole growth,
and the apparent BHAR-SFR relation is largely a secondary effect
due to the star-forming main sequence. 
Among our sources, massive 
galaxies ($M_* \gtrsim 10^{10} M_{\sun}$) 
have significantly higher BHAR/SFR ratios than 
less-massive galaxies, indicating 
the former have higher black-hole 
fueling efficiency and/or higher SMBH occupation 
fraction than the latter.
Our results can naturally explain the observed 
proportionality between $M_{\rm BH}$\ and $M_*$\ 
for local giant ellipticals, and suggest  
their $M_{\rm BH}/M_*$\ is higher than 
that of local star-forming galaxies. 
Among local star-forming galaxies, massive systems 
might have higher $M_{\rm BH}/M_*$\ compared to dwarfs.
\end{abstract}


\section{Introduction}\label{sec:intro}
The origin of the likely coevolution between 
supermassive black holes (SMBHs) and their 
host galaxies remains a fundamental question
\citep[e.g.,][]{hopkins08, marulli08, fabian12, kormendy13}. 
Observations reveal a linear correlation between 
star formation rate (SFR) and 
sample-averaged black-hole accretion 
rate (\bharm) for star-forming 
galaxies \citep[e.g.,][C13 hereafter]{chen13}. 
Also, \xray-selected active galactic nuclei (AGNs) 
preferentially reside in star-forming 
rather than quiescent galaxies
for samples with matched stellar mass  
\citep[$M_*$; e.g.,][]{rosario13}, 
and optically selected luminous quasars 
tend to be hosted by strongly star-forming systems 
\citep[e.g.,][]{harris16, netzer16}.
However, the sample-averaged SFR (\sfrm) of the host 
galaxies of \xray\ AGNs do not show 
a significant dependence on the BHAR 
in the regimes of low and moderate AGN luminosity, 
while the potential existence of a 
positive SFR-BHAR relation 
at high luminosities is still debatable
\citep[e.g.,][]{harrison12, rosario12, barger15, 
stanley15}. 

To reconcile the apparent discrepancy, 
\citet[][H14 hereafter]{hickox14} proposed a 
model in which the long-term 
(\hbox{$\sim 100$~Myr}) average BHAR 
traces SFR linearly, but AGN variability hides 
the BHAR-SFR relation for individual X-ray AGNs
(also see \citealt{rosario13});
SFRs are stable over timescales 
$\gtrsim$~100~Myr, while AGNs are variable over 
much shorter timescales.
This simple scenario reasonably explains 
observations, including both the linear 
\bharm-SFR relation for star-forming galaxies 
and the generally flat \sfrm-BHAR relation for 
\xray-selected AGNs. 

The H14 model requires strong AGN variability 
(by a factor of $\gg 10$) on timescales of 
$\lesssim 10^{7}$~yr to be commonplace.
Although variability studies on the longest 
available timescales ($\lesssim 10$~yr, rest-frame) 
do not directly reveal the prevalence of such 
variability \citep[e.g.,][]{shemmer14, yang16}, its 
occurrence on timescales of $10^2-10^7$~yr is 
plausible from both observational and theoretical 
points of view \citep[e.g.,][]{martini03, novak11}.
In fact, some observational evidence 
suggests the typical AGN-phase time scale is 
$\sim 10^5$~yr, as expected from the chaotic-accretion 
scenario \citep[e.g.,][]{king15b, schawinski15}.
Due to the potential existence 
of such strong variability,
the BHAR derived from direct \xray\ observations
of individual AGNs might not be a reliable 
indicator of long-term average SMBH growth rate. 
On the other hand, \bharm, the average BHAR over
a sample of galaxies, serves as a proxy 
for typical long-term average BHAR of 
the sample (e.g., C13 and H14). 
Therefore, \bharm\ provides a useful tool to 
study SMBH-galaxy coevolution.

Another major motivation of the 
H14 model is that, in the local elliptical galaxies, 
the mass of SMBHs (\mbh) is roughly proportional to 
the bulge $M_*$\ (equivalent to host-galaxy 
$M_*$\ for ellipticals; see \citealt{kormendy13}
for a review). If the long-term average BHAR is proportional 
to SFR for all galaxies, then a natural consequence is 
that \mbh\ correlates with $M_*$\ linearly, as long as
the accreted mass dominates over the mass of
SMBH seeds \citep[e.g.,][]{volonteri10}. 
However, hints have been found of spiral and dwarf 
galaxies hosting undermassive SMBHs relative 
to the \mbh-$M_*$\ relation derived from 
ellipticals, although large uncertainties exist 
\citep[e.g.,][]{greene10, miller15, 
reines15, trump15, greene16}. 
This behavior is not expected from the 
H14 model which assumes SFR is the only factor 
determining long-term average BHAR.
Also, simulations indicate that the 
apparent discrepancy between the \bharm-SFR and 
the \sfrm-BHAR relations can be produced by the 
effect of binning on the intrinsic bivariate 
relationship between BHAR and SFR 
\citep[e.g.,][]{volonteri15}, 
regardless of whether the intrinsic shape of 
this distribution is produced by an intrinsic 
long-term BHAR-SFR relation as proposed by H14.

Observations show that the fraction 
of AGNs above a given luminosity 
threshold rises steeply toward massive galaxies 
\citep[e.g.,][]{xue10, aird12, bongiorno12, mullaney12b}.
Furthermore, for $M_*$-matched samples, the 
fraction of galaxies hosting AGNs 
appears to have no dependence on 
host-galaxy colors 
\citep[e.g.,][]{silverman09, pierce10, xue10}. 
However, apparent 
galaxy colors might be a poor indicator of 
SFR, as high-SFR galaxies might appear red due
to significant dust obscuration 
(e.g., \citealt{whitaker12, rosario13}; and 
references therein). 
Therefore, it is still not clear whether $M_*$\ 
or SFR is the dominant factor correlated with 
black-hole accretion. 


With the advent of deep ultraviolet-to-infrared 
(UV-to-IR) observations from surveys such as CANDELS 
\citep{grogin11, koekemoer11}, it has 
become possible to estimate reliably $M_*$\ and SFR 
for the majority population of galaxies with
acceptable uncertainties ($\lesssim 0.1$\ and $0.2$~dex
for $M_*$\ and SFR, respectively; see 
Sec.~\ref{sec:sfr}). The uncertainties on $M_*$\ 
and SFR are small compared to the parameter ranges 
probed (both $\approx 3$ orders of magnitude), and thus 
are acceptable for our analyses.
The 7~Ms \chandra\ Deep Field-South (\cdfs, covering 
the whole \goodss\ region; see 
\citealt{luo17}, L17 hereafter) 
\xray\ survey has achieved unprecedented sensitivities, 
allowing the derivation of accurate \bharm\ values for 
the galaxies in \goodss.   
In this paper, we evaluate the dependence of \bharm\ on 
both SFR and $M_*$\ for galaxies in \goodss. Also, we study the 
efficiency of SMBH growth compared to SFR for galaxies 
of different $M_*$. 

The paper is structured as follows. We describe the 
sample selection and measurements of SFR, $M_*$, and 
BHAR\ in Sec.~\ref{sec:analyses}. 
In Sec.~\ref{sec:res}, we present the results of our 
analyses. We discuss scientific implications of our 
results in Sec.~\ref{sec:disc}. 
We summarize our results in Sec.~\ref{sec:sum}.

Throughout this paper, we assume a cosmology 
with $H_0=70$~km~s$^{-1}$~Mpc$^{-1}$, $\Omega_M=0.3$, 
and $\Omega_{\Lambda}=0.7$, and a Chabrier
initial mass function \citep[][]{chabrier03}.
Quoted uncertainties are at the $1\sigma$\ (68\%)
confidence level, unless otherwise stated. 
SFR and BHAR are in units of $M_\sun$~yr$^{-1}$, 
and $M_*$\ is in units of $M_\sun$, unless 
otherwise stated. 

\section{Data Analyses}\label{sec:analyses}
\subsection{Sample Selection}\label{sec:samp}
We first select all galaxies with $0.5\leq z<2.0$\ and 
$\mathrm{F160W} <28$\ in the \goodss\ catalog 
\citep[][S15 hereafter]{guo13, 
santini15}.\footnote{$\mathrm{F160W}=28$\ is 
approximately the 5$\sigma$\ limiting magnitude of the 
\goodss\ catalog \citep[][]{guo13}.}
We do not include sources beyond $z=2$, because $M_*$\ and 
SFR values estimated from SED fitting (Sec.~\ref{sec:sfr}) 
suffer from potential biases in that redshift regime 
(e.g., \citealt{wuyts11} and S15).
The SEDs of broad-line AGNs often 
have significant accretion-disk emission 
besides their starlight. Their SFR and $M_*$\
measurements from SED fitting have potential large 
uncertainties \citep[e.g.,][]{bongiorno12, sun15}. 
Thus, we exclude the 19 broad-line AGNs reported in the literature 
\citep[e.g.,][]{mignoli05, ravikumar07, silverman10}, and discuss 
the effects of their exclusion in Sec.~\ref{sec:missed_bhar}.
Applying these criteria, we select 18221 sources 
(see Tab.~\ref{tab:src_num}), 
of which 1305 have secure spectroscopic 
redshifts, and the rest have high-quality photometric 
redshifts based on up to 17 bands from the UV to IR 
(S15). Compared to spectroscopic 
redshifts when available, the photometric redshifts 
have median uncertainty 
$|z_{\rm phot}-z_{\rm spec}|/(1+z_{\rm spec}) \approx 2\%$\
with an outlier (uncertainty $>15\%$) fraction of 3\%.


\begin{table*}
\begin{center}
\caption{Numbers of Sources in Different Samples}
\label{tab:src_num}
\begin{tabular}{crrrc}\hline\hline
Sample & Low-$z$ & High-$z$ & Total & Fig(s). \\ 
(1) & (2) & (3) & (4) & (5) \\ \hline
All & 10057 & 8164 & 18221 & \ref{fig:Lx_vs_z} 
			     and \ref{fig:m_vs_sfr} \\ 
$0.1 \leq \mathrm{SFR} < 100\ M_{\sun}$~yr$^{-1}$\ (A)
		& 6384 & 7541 & 13925 & 
		\ref{fig:Lx_vs_SFR} \\ 
$10^{8} \leq M_* < 10^{11}\ M_{\sun}$\ (B) 
		& 6445 & 6669 & 13114 & 
		\ref{fig:Lx_vs_M} and \ref{fig:ratio_vs_M} \\ 
(A) $\land$\ (B) & 5224 & 6236 & 11460 & 
		\ref{fig:BHAR_vs_M_and_SFR} \\ \hline
\end{tabular}
\end{center}
{\sc Note.} --- Columen (1): sample definition. 
Columns (2) and (3): number of sources with $0.5 \leq z < 1.3$\ and 
	     $1.3 \leq z < 2.0$, respectively. 
Column (4): total number of sources in both redshift ranges. 
Column (5): relevant figure(s).\\
\end{table*}

\subsection{Stellar Mass and Star Formation Rate}\label{sec:sfr}
\begin{figure}[htb]
\includegraphics[width=\linewidth]{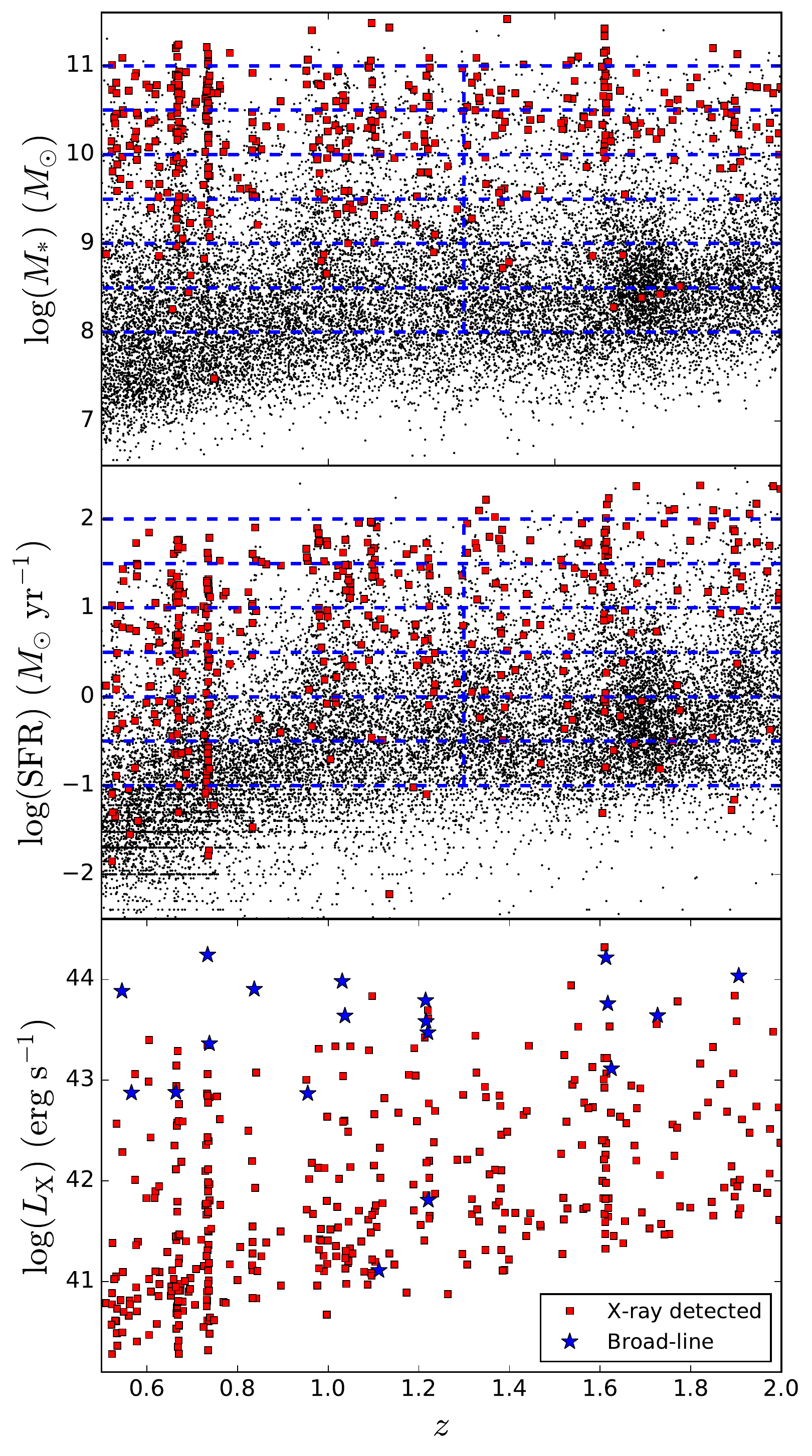}\\
\caption{The top (middle) panel shows $M_*$\ (SFR) 
vs.\ redshift for all our sources.
The red squares indicate \xray\ detected sources.
The blue dashed lines indicate our binning grids in 
Figs.~\ref{fig:Lx_vs_SFR}, \ref{fig:Lx_vs_M}, 
and \ref{fig:ratio_vs_M}. 
The bottom panel presents absorption-corrected 
\lx\ (see Sec.~\ref{sec:bhar}) as a 
function of redshift for \xray\ detected sources. 
The blue stars indicate broad-line AGNs
that are excluded from our sample.
The overdensities at $z\approx 0.7$\ and 1.6\ are 
likely due to cosmic variance (e.g., 
\citealt{silverman10,finoguenov15}; L17).
}
\label{fig:Lx_vs_z}
\end{figure}

\begin{figure}[htb]
\includegraphics[width=\linewidth]{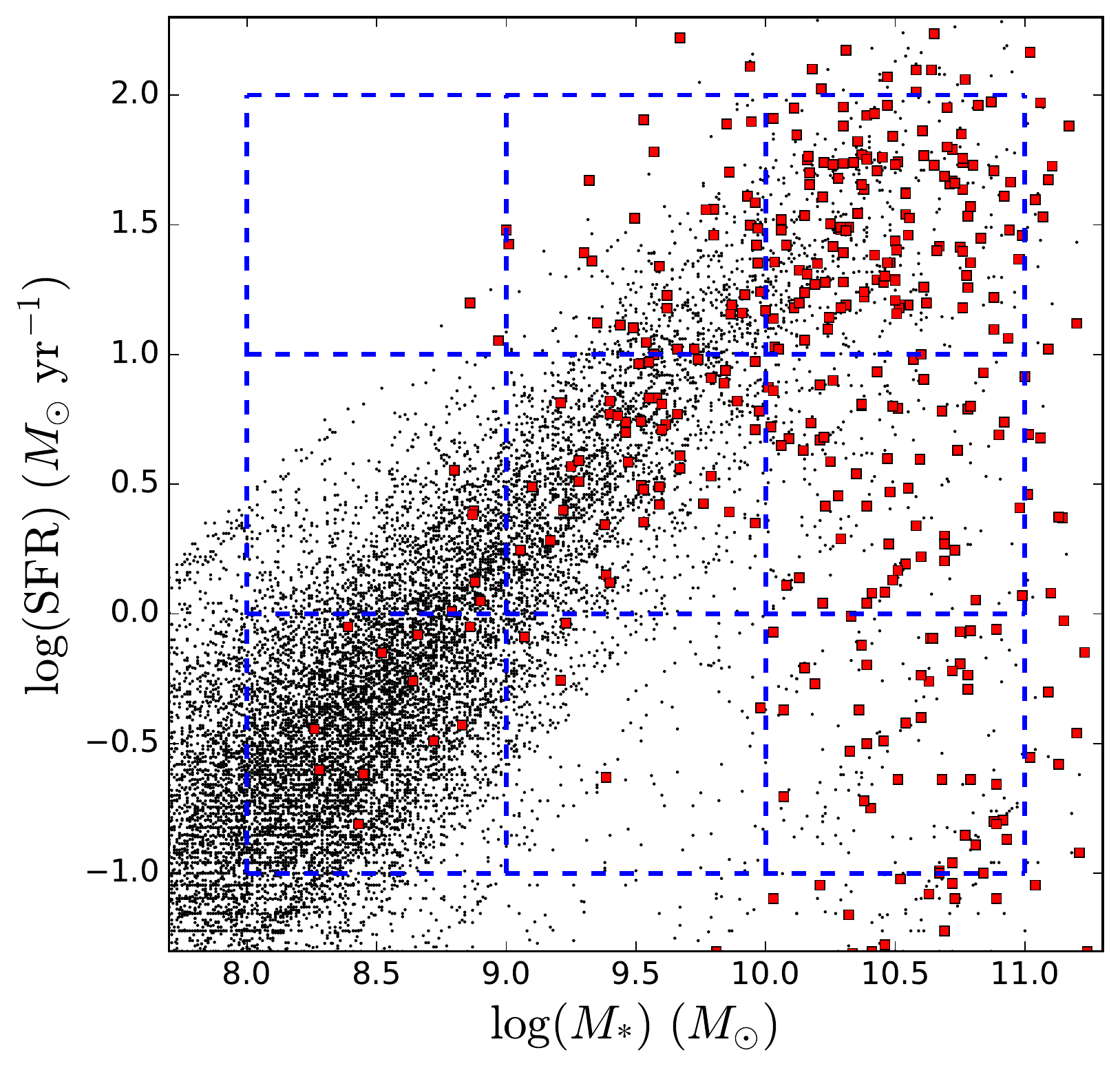}\\
\caption{SFR vs.\ $M_*$\ for all our sources. 
The red squares indicate \xray\ detected sources.
The blue dashed lines indicate our binning grids 
in Fig.~\ref{fig:BHAR_vs_M_and_SFR}.}
\label{fig:m_vs_sfr}
\end{figure}

We collect the $M_*$\ and SFR values for our sources from S15, 
who presented SED-fitting results of several 
independent teams. 
We adopt the median values of $M_*$\ and SFR
from the five available teams (i.e., 
labeled as 2a$_\tau$, 6a$_\tau$, 11a$_\tau$, 13a$_\tau$, 
and 14a in S15). All five teams employed stellar templates 
from \cite{bruzual03} and a Chabrier IMF when performing 
the SED fitting. Teams 2a$_\tau$, 6a$_\tau$, 
11a$_\tau$, and 13a$_\tau$\ assumed an exponentially 
declining star formation history (SFH); 14a assumed a 
more flexible SFH (see S15).
Teams 2a$_\tau$, 6a$_\tau$, 11a$_\tau$, and 14a 
adopted Calzetti extinction law \citep{calzetti00}; 
13a$_\tau$\ adopted a combination of Calzetti and SMC
extinction laws.
The $M_*$\ and SFR values estimated by the five teams 
agree well; their typical deviations 
from the adopted medians are $\lesssim 0.1$~dex\
and $\lesssim 0.2$~dex, respectively.
These values largely represent uncertainties 
arising from SED fitting, and additional systematic errors 
(from, e.g., IMF and SFH assumptions) likely exist. 
However, such systematic uncertainties should not affect our 
conclusions qualitatively (see Sec.~\ref{sec:res}).
Fig.~\ref{fig:Lx_vs_z} shows $M_*$\ and SFR
as functions of redshift, and Fig.~\ref{fig:m_vs_sfr}
shows the $M_*$-SFR plane for all our sources. 
From Figs.~\ref{fig:Lx_vs_z} (top panel) and 
\ref{fig:m_vs_sfr}, \xray\ detected 
sources\footnote{These are sources presented in 
the 7~Ms \cdfs\ main catalog (Sec.~\ref{sec:bhar}), 
formally defined with ``binomial no-source probability'' 
$P_{\rm B}<0.007$ (see L17).
}
are preferentially found among massive galaxies 
(also see, e.g., \citealt{xue10, aird12}).
Our sample is roughly complete for galaxies with 
$M_* \gtrsim 10^{8}\ M_{\sun}$\ and $\mathrm{SFR} \gtrsim 0.1\ 
M_{\sun}\ \mathrm{yr^{-1}}$\ (i.e., the $M_*$\ and SFR regimes
mostly probed by our analyses; see Sec.~\ref{sec:res}).
A more detailed discussion of completeness is presented in 
Sec.~\ref{sec:complete}.

The rest-frame UV-to-near-IR SEDs ($\approx 0.2-4$~$\mu$m, 
similar to the wavelength range used 
by S15 to derive $M_*$\ and SFR) of \xray-selected AGNs
in the \cdfs\ are usually dominated by stellar light
(see Fig.~9 of \citealt{luo10}; also see, e.g.,  
\citealt{xue10, brandt15}). In addition, we have excluded 
broad-line AGNs (Sec.~\ref{sec:samp}), 
because their $M_*$\ and SFR measurements from 
SED fitting could be overestimated 
\citep[e.g.,][]{ciesla15}.
Therefore, the CANDELS $M_*$\ and SFR should not 
have significant biases due to AGN activity.
Also, we confirm that AGNs do not have biased 
SED-based SFRs in comparison with SFRs based on 
far-IR (FIR) photometry (see below).  

As demonstrated by S15, the CANDELS $M_*$\ values 
generally have high quality. To evaluate the accuracy 
of CANDELS SED-based SFRs, which ultimately come from 
dust-corrected UV luminosities, we compare them with 
those obtained from FIR photometry. 
We match our sources with the 
\herschel/PACS catalog of the PACS Evolutionary Probe 
(PEP; e.g., \citealt{lutz11, magnelli13}) 
survey using a 2$''$\ matching radius.\footnote{We use the 
\hbox{24~$\mu$m}-prior PEP catalog due 
to its good positional accuracy.}
We follow the method of C13 to derive 
FIR-based SFRs for the matched sources.
Briefly, we convert the PACS band flux to 
total IR luminosity according to the star-forming 
galaxy spectral template from 
\cite{kirkpatrick12},\footnote{\cite{kirkpatrick12} 
present two templates at $z\sim 1$\ and $z\sim 2$,
respectively. Here, we use the $z\sim 1$\ template 
following C13, although the $z\sim 2$\ template 
leads to almost the same results.}
and then we scale the IR luminosity to SFR 
following the relation from \cite{kennicutt98}
(modified for our Chabrier IMF; see C13).
We use the PACS 100~$\mu$m band for sources at 
$0.5\leq z < 1.3$\ and the 160~$\mu$m band at 
$1.3\leq z < 2.0$, requiring photometric 
$\mathrm{S/N}>5$. Using different bands 
for different redshift ranges is to sample 
the SED peak 
(\hbox{$\approx 50-80$~$\mu $m}, rest-frame; e.g., 
\citealt{kirkpatrick12})
of cold dust emission, a good proxy for star formation.
The FIR-based SFR estimation assumes that dust absorbs 
most UV photons and reemits IR radiation. 
It is thus robust for galaxies with relatively 
high SFR ($\gtrsim 1\ M_{\sun}$~yr$^{-1}$) where 
dust is often abundant 
\citep[e.g.,][]{calzetti10, kennicutt12}, 
and this is indeed our case, since low-SFR galaxies 
within our redshift range ($0.5\leq z < 2.0$)
are usually not detected by \herschel.

\begin{figure}[htb]
\includegraphics[width=\linewidth]{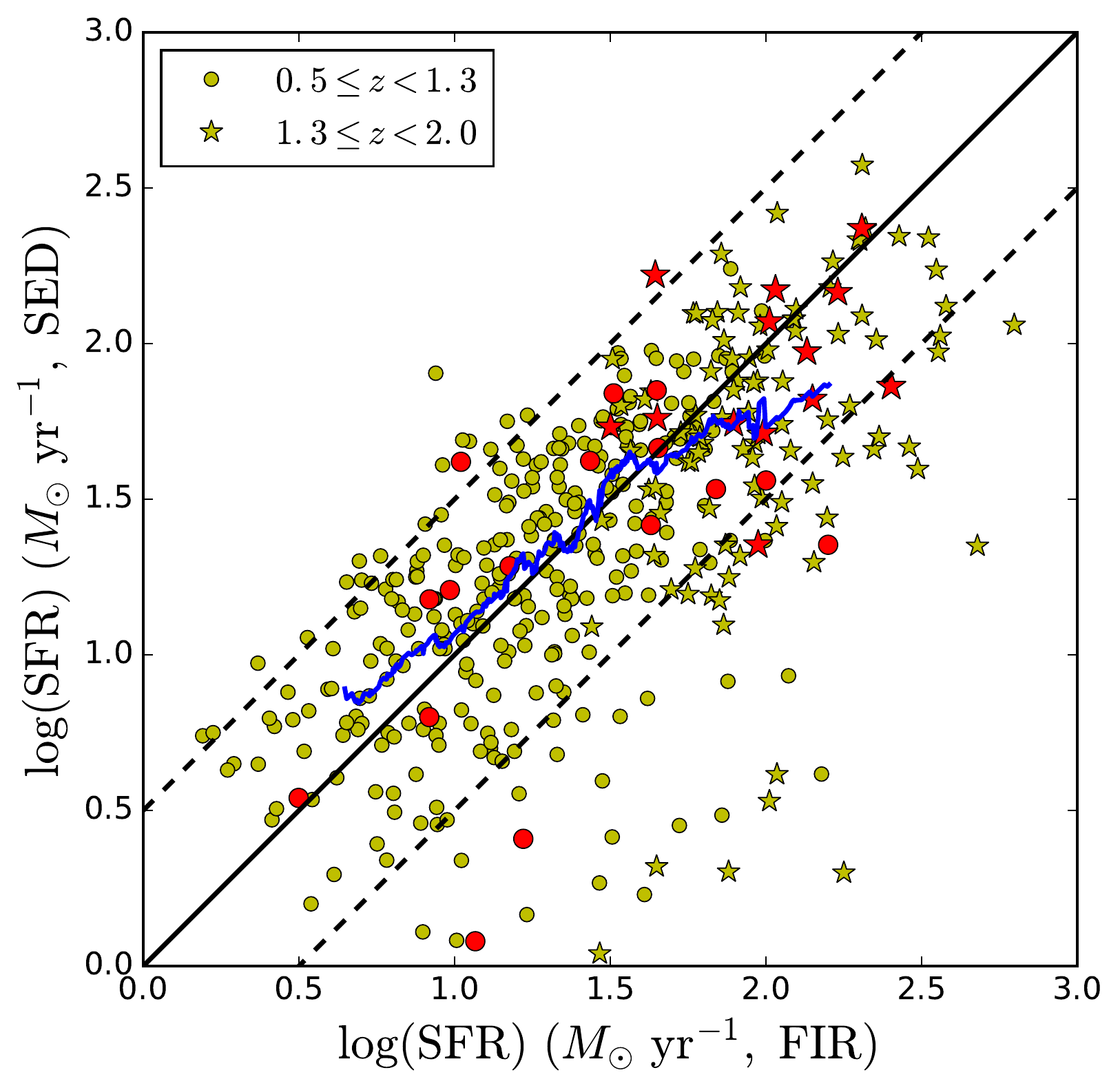}\\
\caption{Comparison between SFRs derived from SED
fitting and those derived from FIR photometry. 
Circles and stars indicate sources in the 
redshift ranges of $0.5\leq z <1.3$\ and 
$1.3\leq z <2.0$, respectively. Their FIR-based 
SFRs are derived from 100~$\mu $m and 160~$\mu $m
observations with \herschel/PACS, respectively
(see Sec.~\ref{sec:sfr}).
Red symbols indicate AGNs selected 
by their large \xray\ luminosities 
\hbox{(\lx$>3\times 10^{42}$~erg~s$^{-1}$; 
see e.g., L17)}.
The black solid line indicates 1:1 relation
between two SFR measurements; the dashed lines 
indicate 0.5-dex offsets.
Our adopted SFRs (from SED fitting) generally agree 
with those derived from FIR photometry. 
The blue solid curve indicates running median 
SFR offsets from bins of 50 sources.
For \xray-luminous AGNs, the SED-based SFRs do not 
have significant systematic differences relative to 
FIR-based SFRs.
}
\label{fig:SFR_vs_SFR}
\end{figure}

Fig.~\ref{fig:SFR_vs_SFR} compares 
CANDELS SED-based SFRs with 
FIR-based SFRs for \herschel-detected galaxies 
that probe the SFR range of 
$\approx 10^{0.5}-10^{2.5}\ M_{\sun}$~yr$^{-1}$. 
CANDELS SFRs generally agree well 
with FIR-based SFRs without significant systematic 
bias: the median offset between them is 
0.01~dex. 
The median offset depends on SFR (the blue 
solid curve in Fig.~\ref{fig:SFR_vs_SFR}). 
The offset is small ($\lesssim 0.2$~dex)
at low and intermediate SFR, but becomes $\approx 0.3$~dex 
at $\rm SFR \sim 100$~$M_\sun$~yr$^{-1}$.
The median offsets for the low-$z$\ and high-$z$\ bins 
are 0.07 and 0.16~dex, respectively. 
A similar systematic offset in the high-SFR (also high-$z$) 
regime is also found by \cite{wuyts11}, possibly
because dust correction cannot fully recover the 
intrinsic UV luminosity when the obscuration is 
very strong. 
Nevertheless, the systematic errors are smaller 
than our bin width of SFR (i.e., 0.5 and 1~dex; 
see Sec.~\ref{sec:res}), and thus should not affect 
our results significantly.

For most sources (80\%), 
the SFRs derived by the two methods 
agree within 0.5~dex 
(the dashed lines in Fig.~\ref{fig:SFR_vs_SFR}). 
The outliers (20\%) tend to have FIR-based SFRs 
higher than SED-based SFRs, possibly because 
the FIR sample is flux-limited and 
FIR-luminous outliers are more likely to be 
detected by \herschel
(see also, e.g., \citealt{azadi15}).
There are 6\% extreme outliers with
SFR offsets larger than 1~dex. 
For these sources, the SFR measurements from 
different SED-fitting teams are often inconsistent; 
their typical deviations 
from the adopted medians are $\approx 0.3-1$~dex,
significantly larger than those for the whole sample 
($\lesssim 0.2$~dex). 
The likely failure of SED fitting might be caused 
by inappropriate model assumptions in SED fitting, 
e.g., SFH and extinction law. 
False matches between the CANDELS and FIR catalogs 
might also be responsible for some of the extreme 
outliers. 
Nevertheless, none of these extreme outliers
has high \xray\ luminosities 
($> 3\times 10^{42}$~erg~s$^{-1}$), and their 
large SFR uncertainties are unlikely to affect
our results qualitatively.
Notably, for AGN-dominated \xray\ sources  
with \hbox{\lx$>3\times 10^{42}$~erg~s$^{-1}$}
(see L17; red symbols in Fig.~\ref{fig:SFR_vs_SFR}), 
the CANDELS SFRs do not show a significant systematic 
bias relative to SFRs derived from FIR photometry:
the median offset is 0.01~dex, similar as the 
value for all \herschel-detected galaxies above.
Therefore, the SED-based SFRs are reliable for 
galaxies with 
$\mathrm{SFR} \gtrsim 10^{0.5}\ M_{\sun}$~yr$^{-1}$.
In the lower-SFR regimes, corrections for dust 
extinction are generally low or moderate, 
and SED-based SFRs are generally reliable 
\citep[e.g.,][]{wuyts11, kennicutt12}.

\subsection{Black-Hole Accretion Rate}\label{sec:bhar}
We use \xray\ observations to derive \bharm\
for our sources. We match our 18221 galaxies with the 7~Ms 
main source catalog for the \cdfs\ (L17) 
using a 0.5$''$\ matching 
radius, and 395 \xray\ sources are matched.\footnote{We 
use the positions of the L17 optical/near-infrared 
counterparts rather than the \xray\ positions, 
because the former are more accurate. In the \goodss\ 
field, there are 704 out of the 1008 \xray\ sources in
the L17 catalog, and most of them 
($674/704 = 96\%$) have CANDELS counterparts. The 
remaining 30 sources without counterparts might be 
faint (F160W band) sources not detected by CANDELS, 
nearby off-nuclear sources (e.g., ultraluminous \xray\ 
sources), or false \xray\ detections (L17). 
395 of the 674 \xray\ sources with CANDELS counterparts 
are galaxies in our redshift range ($0.5 \leq z < 2.0$)
without broad lines reported in their spectra. 
}
A total of 259 of the 395 \xray-detected sources have 
spectroscopic redshifts. 
The host-galaxy properties ($z$, $M_*$, and SFR) 
of these \xray\ detected sources are shown in 
Figs.~\ref{fig:Lx_vs_z} and \ref{fig:m_vs_sfr}.

We fit their unbinned \xray\ spectra 
(observed-frame \hbox{0.5--7 keV}) 
with the Cash statistic \citep{cash79}.
We perform the fitting with a standard absorbed power-law 
model (i.e., $wabs\times zwabs \times powerlaw$\ in XSPEC; 
see \citealt{arnaud96} for a description of XSPEC) to 
recover their absorption-corrected \xray\ luminosities
(\lx, rest-frame \hbox{2--10~keV}; e.g., \citealt{yang16}). 
The $wabs$\ component accounts for Galactic absorption with 
absorption column density (\nh) set to 
$8.8\times 10^{19}$~cm$^{-2}$\ \citep{stark92}. 
The $zwabs$\ component models intrinsic 
absorption (i.e., $wabs$\ at redshift $z$).
The normalization of $powerlaw$,
intrinsic photon index, and intrinsic \nh\ are free 
parameters in the fitting. 
The allowed ranges of photon index and \nh\ are set to 
$1.4-2.2$\ and \hbox{$10^{19}-10^{24}$~cm$^{-2}$}, 
respectively. 
We then obtain the \lx\ with XSPEC from the best-fit 
model parameters. The best-fit \lx\ as a function 
of redshift is shown in Fig.~\ref{fig:Lx_vs_z} (bottom). 
Thanks to the great sensitivity of the 7~Ms \cdfs\
survey, sources with low \xray\ luminosities 
($\lesssim 3 \times 10^{42}$~erg~s$^{-1}$) can be detected 
up to $z=2$. \xray\ emission at this low level might not be  
dominated by AGNs, but could also originate from stellar 
processes.

Our absorbed power-law model is appropriate for moderately 
obscured or unobscured AGNs; it might result in 
unreliable \lx\ for Compton-thick AGNs (CTK AGNs; i.e., 
$N_{\rm H} \gtrsim 10^{24}$~cm$^{-2}$). 
However, it is generally challenging to identify 
\textit{bona-fide}
CTK AGNs among the \xray\ detected sources 
in deep fields such as \cdfs\ due 
to the limited numbers of counts available, 
and there is no strong evidence suggesting that
CTK AGNs are the dominant population 
(e.g., \citealt{alexander13, brightman14}; Sec.~3.3
of \citealt{brandt15}).
We have tested a basic CTK model, 
$wabs\times zwabs \times pexmon$, on our \xray\
detected sources (see \citealt{magdziarz95} and 
\citealt{nandra07} for $pexmon$; see App.~B of \citealt{yang16}
for model parameter settings). Only 4\% of sources show 
statistically significant improvements in fitting 
compared to our adopted 
$wabs\times zwabs \times powerlaw$\ model, 
where we use the Akaike information criterion 
to infer fitting improvement (see \citealt{yang16} 
for details). Thus, CTK sources are not 
likely to be the dominant population in our sample.
Depending on model assumptions (e.g., obscuration 
geometry and viewing angle),
the CTK models available often have 
large uncertainties in the derived physical 
parameters, e.g., \lx\ and \nh\  
\citep[see, e.g.,][]{murphy09}. 
Therefore, we do not adopt the spectral-fitting
results from the CTK models.
A discussion on the effects of \xray\ non-detected 
CTK sources is presented in Sec.~\ref{sec:missed_bhar}.

For sources without \xray\ detections in each sample in 
Sec.~\ref{sec:res}, we perform a stacking analysis 
to derive their total \lx\ following the procedures of
\cite{vito16}.  
Briefly, we convert the stacked total count rate 
of the soft band\footnote{Observed-frame 
\hbox{0.5--2 keV}. The soft band has larger collecting 
area and lower background, than the hard band 
(\hbox{observed-frame 2--7 keV}), e.g., the expected count
rate in the soft band is $\approx 2$\ times larger 
than that in the hard band for a $\Gamma=1.8$\ power-law 
spectrum. We have tested stacking the count rate 
in the hard band. However, the resulting S/N is generally 
much weaker than that from soft-band stacking, and 
the hard-band stacked count rates are consistent 
with zero in many cases. 
} to total \lx\
($L_{\rm X, stack}$) of the stacked sources 
assuming their median redshift\footnote{The median 
and mean redshifts for our samples (Sec.~\ref{sec:res}) 
are similar, and they only differ by $\lesssim 0.02$.} 
and a power-law spectrum with an effective 
photon index of $\Gamma=1.8$. 
By setting $\Gamma=1.8$, we assume the stacked 
signals are mainly from \xray\ emission of \xray\ 
binaries (XRBs) and/or AGNs with low obscuration of 
$N_{\rm H} \lesssim 10^{21.5}$~cm$^{-1}$\ (see Sec.~6.1
of \citealt{lehmer16} for a detailed discussion).
Indeed, our stacked \xray\ fluxes are similar to 
those expected from XRBs (see 
Sec.~\ref{sec:bhar_vs_sfr} and \ref{sec:bhar_vs_m}).
Even if we adopt a very flat spectral shape of 
$\Gamma=1$\ (i.e., assume all the stacked \xray\ 
signals are entirely caused by moderately obscured 
AGNs with $N_{\rm H} \sim 10^{22.5}$~cm$^{-2}$),
the resulting $L_{\rm X, stack}$\ will be only 
$\sim 0.3$~dex higher, unlikely to have a large 
impact on our qualitative results 
(see Sec.~\ref{sec:res}).
We discuss the potential effects of heavily 
obscured AGNs that might not be included in our 
analyses in Sec.~\ref{sec:missed_bhar}. 
Following \cite{vito16}, we exclude sources 
that are at off-axis angles greater than $7.8'$\ or 
close to \xray\ detected sources (see \citealt{vito16} 
for the specific criteria) in the stacking analyses. 
Those excluded sources represent a real population 
of galaxies, and thus their contribution to \bharm\
should be included in our analyses. 
We account for the excluded sources ($\lesssim 25\%$\ 
of the \xray-undetected sources)
in each sample by assuming that their mean 
\lx\ is the same as that of the stacked sources, and scale 
the $L_{\rm X, stack}$\ by multiplying 
$N_{\rm non}/N_{\rm stack}$\
to obtain the total \lx\ of all 
\xray-undetected sources (Eq.~\ref{equ:lx}). 

XRBs and other stellar processes 
in galaxies also contribute to the observed \lx, and thus we 
need to subtract their contribution (\lxg).\footnote{\xray\
emission from XRBs usually dominates over that from other 
stellar processes. We assume that all non-AGN \hbox{X-rays}
are XRB contributions.} The \lxg\ is 
estimated as \lxg$=\alpha M_* + \beta \rm{SFR}$, where
$\alpha$\ and $\beta$\ are 
coefficients as functions of redshift.
We adopted the redshift-dependent $\alpha$\ and $\beta$\ values 
from model 269 of \cite{fragos13} 
[corrected to our Chabrier IMF following the 
prescriptions from 
\cite{longhetti09} and \cite{madau14}] 
that is preferred by the observations of 
\cite{lehmer16}.\footnote{We have also tested 
a simpler \lxg\ model \citep{ranalli03} in which 
$\alpha$\ is zero and $\beta$\ is a constant
(i.e., not dependent on redshift). 
Our results below only change slightly using this model.}
Our adopted $\alpha$\ and $\beta$\ are also consistent 
with the values from \cite{aird16}: the diffrences 
are $\approx 0.1$~dex for both $\alpha$\ and $\beta$\
in $0.5 \leq z < 2.0$.
The mean AGN \lx\ for each sample is calculated as 
\begin{equation}\label{equ:lx}
\langle L_{\rm X} \rangle = \frac{ (\sum_{\rm detect} L_{\rm X}) +
 		\frac{N_{\rm non}}{N_{\rm stack}} L_{\rm X, stack}  
		-  \sum_{\rm all} L_{\rm X,XRB}}
		{N_{\rm detect}+N_{\rm non}},
\end{equation}
where $N_{\rm detect}$, $N_{\rm non}$, and $N_{\rm stack}$\ 
are the numbers of \xray-detected, undetected, 
and stacked sources in the sample, respectively.
We do not exclude the 18 radio-loud AGNs
identified by \cite{bonzini13} in our analyses,
although excluding them would have only minor effects 
on our results.
Jet-linked \xray\ emission might contribute 
to their \lx, but, at least for the two 
\xray\ brightest 
radio-loud AGNs, detailed studies do not reveal
significant jet-linked \xray\ emission 
\citep[e.g.,][]{iwasawa15}.

We convert \lxm\ to mean BHAR as  
\begin{equation}\label{equ:bhar}
\begin{split}
\langle \mathrm{BHAR} \rangle &= \frac{(1-\epsilon) k_{\rm bol} 
			\langle L_{\rm X} \rangle}{\epsilon c^2} \\
	 &= \frac{3.53 \langle L_{\mathrm X} \rangle}{10^{45}\ \rm{erg~s^{-1}}}
	    M_{\sun}\ \mathrm{yr}^{-1}
\end{split}
\end{equation}
where we assume a constant bolometric correction 
factor of $k_{\rm bol}=22.4$\ 
(the median value for the local AGN sample with 
$L_{\rm X}\approx10^{41}-10^{45}$~erg~s$^{-1}$\ in 
\citealt{vasudevan07})
and a constant mass-energy conversion efficiency 
of $\epsilon=0.1$\ \citep[e.g.,][]{marconi04, davis11}. 
We obtain the $1\sigma$\ confidence interval as the 
range between the 16th and 84th percentiles of the 
bootstrapped \bharm\ distribution. 
To obtain the \bharm\ distribution, we randomly 
resample the sources 1000 times. 
In this routine procedure of bootstrapping, 
each random resampling includes the same number 
of sources as the original sample but allowing 
repetition of sources. We then calculate \bharm\ with 
Eqs.~\ref{equ:lx} and \ref{equ:bhar} for each 
resampling and obtain the \bharm\ distribution.

From Eq.~\ref{equ:bhar}, we obtain \bharm\ 
from \lxm. This is because X-ray emission is 
almost a universal tracer of black-hole accretion 
\citep[e.g.,][]{gibson08}. X-rays are also relatively 
less affected by obscuration compared to the UV/optical 
bands, and suffer minimal starlight dilution 
\citep[e.g.,][]{brandt15}. Nevertheless, there might 
be uncertainties in the conversion factors 
($k_{\rm bol}$\ and $\epsilon$) between \bharm\ and 
\lxm.
The $k_{\rm bol}=22.4$\ and $\epsilon=0.1$\ 
assumptions have been widely adopted in previous 
studies related to black-hole accretion 
(e.g., \citealt{mullaney12} and C13).
We have also tested
applying a luminosity-dependent $k_{\rm bol}$\ 
\citep{hopkins07} for the AGN-dominated sources with 
\lx$>3\times 10^{42}$~erg~s$^{-1}$\ (e.g., L17), and 
our results do not change qualitatively. Applying 
the luminosity-dependent $k_{\rm bol}$\ to 
low-luminosity sources requires careful subtraction
of \lxg\ for each individual source, but this 
correction is 
beyond the scope of our analyses. For simplicity 
and consistency over all sources, 
we adopt a constant $k_{\rm bol}$\ in our analyses.

Since we include all \xray\ 
detected and non-detected sources (Eq.~\ref{equ:lx}), 
we are measuring \bharm\ averaged over all 
galaxies, including systems with both high and low
levels of nuclear activity. 
This is designed to approximate 
the long-term average BHAR for the entire galaxy sample
rather than the instantaneous BHAR for individual AGNs 
(see Sec.~\ref{sec:intro}). 

\section{Results}\label{sec:res}
\subsection{BHAR Dependence on SFR}\label{sec:bhar_vs_sfr}
\begin{figure*}[htb]
\includegraphics[width=0.52\linewidth]{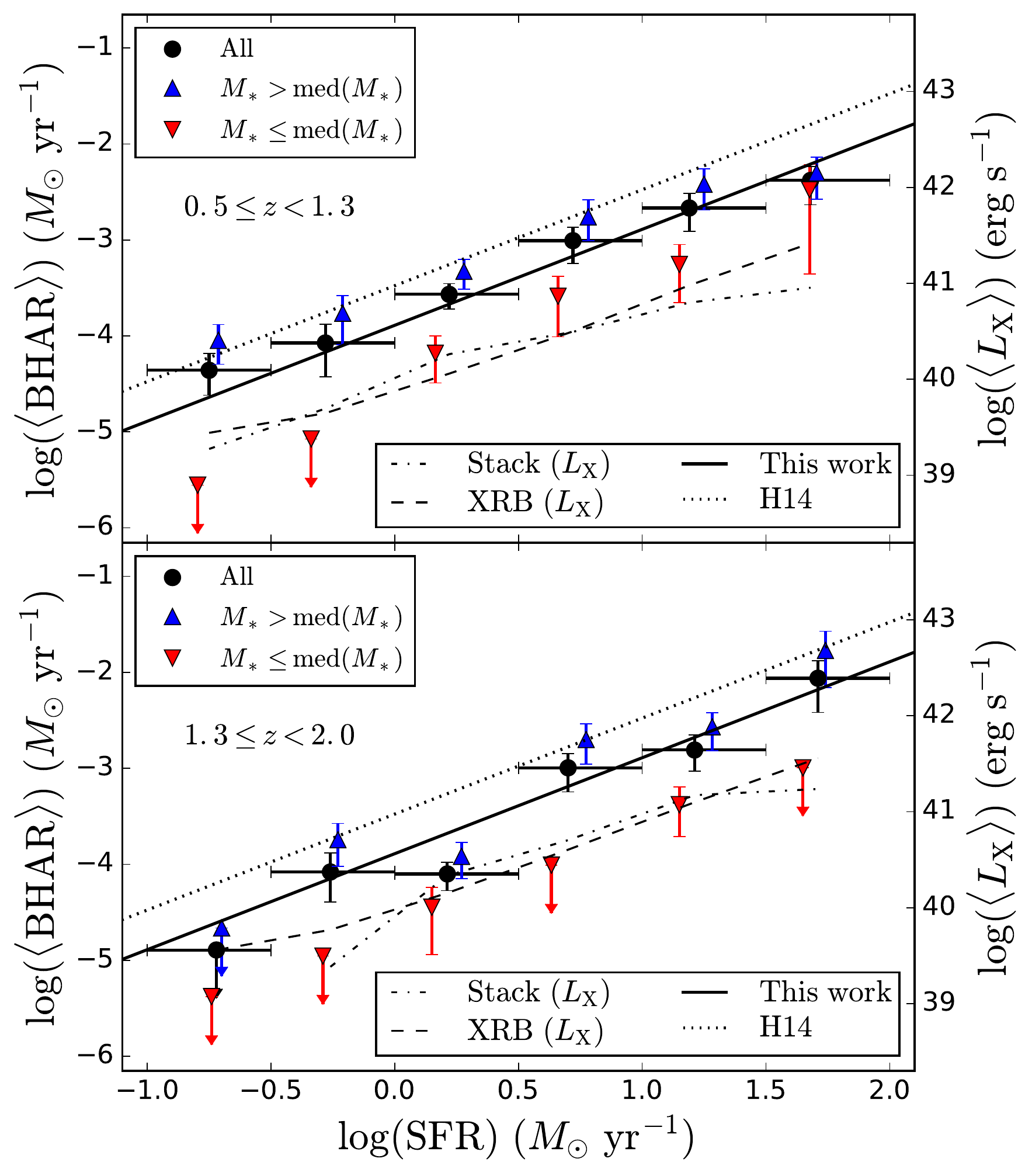}
\includegraphics[width=0.48\linewidth]{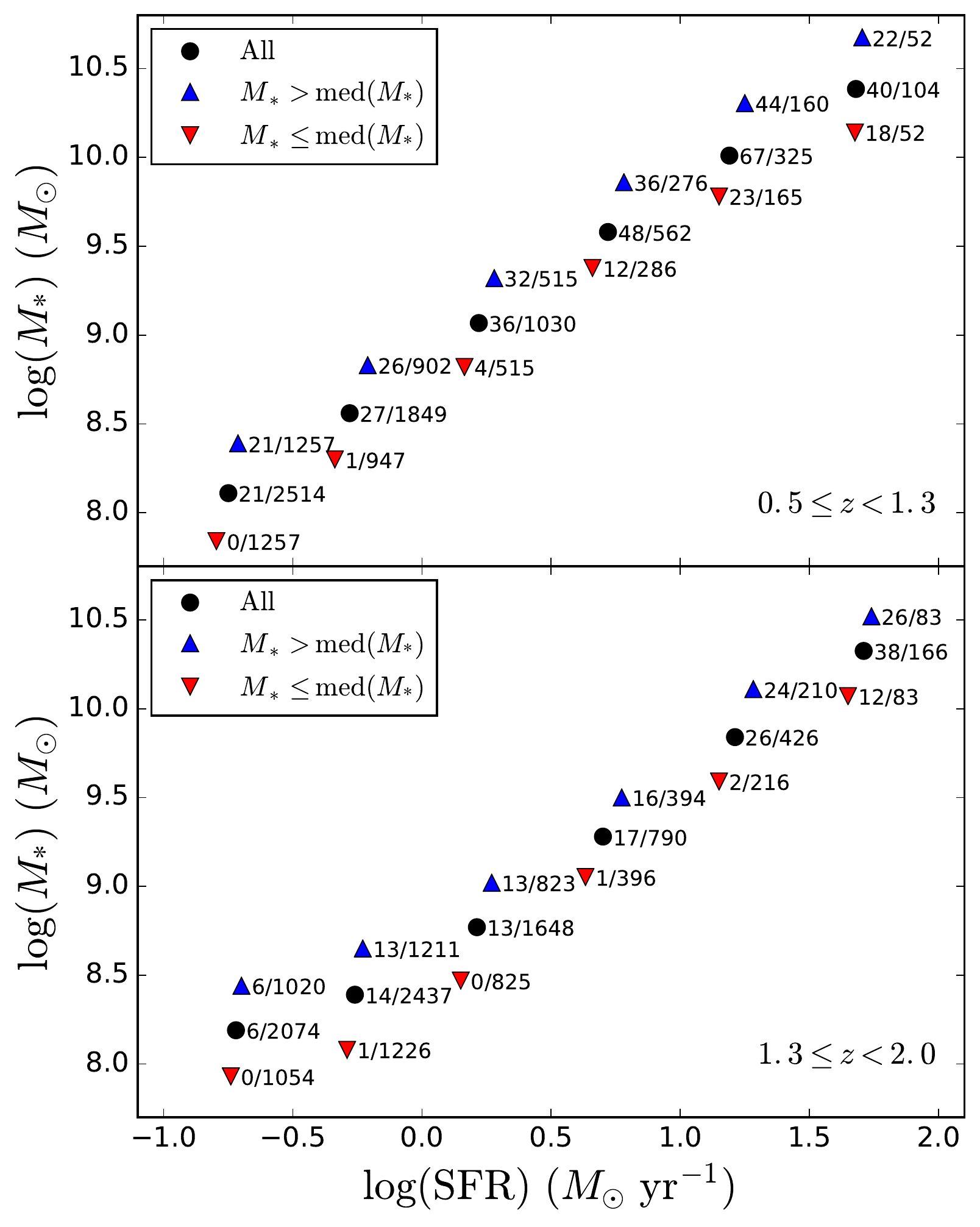}
\caption{Left panels: 
\bharm\ as a function of SFR for different 
redshift ranges. The vertical and horizontal positions of 
the black points indicate \bharm\ and median SFR, 
respectively; the horizontal error bars indicate the bin 
width. 
The blue upward-pointing and 
red downward-pointing triangles indicate 
\bharm\ for two subsamples with $M_*> \med(M_*)$\ and 
$M_*\leq \med(M_*)$, respectively, where $\med(M_*)$\ is the 
median $M_*$\ of sources in each bin 
(Sec.~\ref{sec:bhar_vs_sfr_and_m}). 
The solid black line indicates our best-fit linear 
model with slope fixed at unity;
the dotted black line indicates the model proposed
by H14.
The values of \bharm\ are converted from \lxm, which
is labelled on the right side of each panel.
\lxm\ is derived considering both \xray\ detected and 
undetected (via stacking) sources, and contamination from 
galaxies is subtracted (see Sec.~\ref{sec:bhar}).
All the errors are estimated via bootstrapping 
(1000 simulations).
If the resulting 1$\sigma$\ lower limit of \lxm\
is negative, we use the \lxm\ 
expected from XRB emission as an upper limit. 
The dashed black line indicates the \lxgm\ 
that has been subtracted for each sample.
The dash-dotted line indicates average \xray\ 
luminosities for stacked sources 
(XRB contributions not subtracted),
and it is generally comparable to \lxgm. 
In the lower panel, the lowest-SFR bin has negative 
stacked \lxm\ due to weak \xray\ signals from 
the stacked sources and background fluctuations; 
thus, the dash-dotted line does not 
extend to the bin with the lowest SFR. 
Our results can be fitted acceptably by a linear 
relation between \bharm\ and SFR, but more massive 
galaxies generally have higher \bharm\ at a given
SFR level.
Right panels: the median $M_*$\ corresponding to 
each (sub)sample on the left. The numbers of \xray\ 
detected sources and all sources in each (sub)sample 
are marked on the right side of the corresponding 
point. 
}
\label{fig:Lx_vs_SFR}
\end{figure*}

We bin our sources in six different SFR intervals 
(bin width $=0.5$~dex, see Fig.~\ref{fig:Lx_vs_z}) 
for two different redshift ranges ($0.5 \leq z < 1.3$\
and $1.3 \leq z < 2.0$), and calculate 
the \bharm\ for each of the 12 bins which together
include 13925 sources (Tab.~\ref{tab:src_num})
using Eqs.~\ref{equ:lx} and \ref{equ:bhar}. 
In the following analyses, 
we discard all bins that have fewer than 
100 sources to avoid large statistical fluctuations,
unless otherwise stated.\footnote{Without this 
constraint, we can only extend the SFR and $M_*$\ ranges
by $\approx 0.5$~dex (i.e., one bin; see Figs.~\ref{fig:Lx_vs_SFR}
and \ref{fig:Lx_vs_M}). Such extended samples have very large 
uncertainties on \bharm\ ($\sim 1$~dex or only upper 
limit available) likely caused by statistical fluctuations,
since each sample has only $\lesssim 20$~sources.} 
This selection is the main reason why we cannot probe the 
high-SFR ($\mathrm{SFR} > 100$~$M_{\sun}$~yr$^{-1}$) 
and high-$M_*$\ ($M_* > 10^{11}\ M_{\sun}$) regimes
(Sec.~\ref{sec:bhar_vs_m}).

Fig.~\ref{fig:Lx_vs_SFR} displays the results.  
In general, galaxies with higher SFR have higher 
\bharm. We perform a least-$\chi^2$\ fitting for 
the \bharm-SFR relation with a linear 
model,\footnote{\label{foot:fit} 
We employ the Python code, scipy.optimize.curve\_fit, 
to performing the fitting. 
We use the median SFR of each bin when we perform the 
fitting. We adopt the mean values of $1\sigma$ upper and 
lower uncertainties on \bharm\ in the fitting. 
We do not apply an error to the median 
SFR for simplicity, since the SFR distribution 
in each bin usually has a strong non-Gaussian shape. 
The code estimates the uncertainties on best-fit 
parameters from the covariance matrix. 
Median SFR and mean SFR are very close,
since sources in each bin have similar SFR. 
Using the median or mean does not affect our results 
significantly.
}
and obtain 
\begin{equation}\label{eq:bhar_sfr1}
\log (\langle \mathrm{BHAR} \rangle) 
=(0.93\pm0.08) \log (\mathrm{SFR})
- (3.85\pm0.07) 
\end{equation}
with reduced $\chi^2=1.48$\
which corresponds to a model-rejection 
\hbox{$p$-value} of 15\%. 
The slope is consistent with unity. 
To compare with the H14 model
which assumes \bharm\ is proportional to SFR
($\langle \mathrm{BHAR} \rangle = \mathrm{SFR}/3000$), 
we fix the slope to one and refit the data. 
This fit results in 
\begin{equation}\label{eq:bhar_sfr}
\log (\langle \mathrm{BHAR} \rangle) 
=\log (\langle \mathrm{SFR} \rangle)
- (3.89\pm0.07)
\end{equation}
(the solid lines in
Fig.~\ref{fig:Lx_vs_SFR}) and a reduced 
$\chi^2$\ of 1.40 (\hbox{$p$-$\mathrm{value}=17\%$}).
The best-fit intercept ($-3.89$) 
is \hbox{$\approx 0.4$~dex}
lower than that expected from the H14 model 
(i.e., $-3.48$; shown as the dotted 
lines in Fig.~\ref{fig:Lx_vs_SFR});
possible reasons are explained
in Secs.~\ref{sec:bhar_vs_sfr_and_m} 
and \ref{sec:missed_bhar}.
Our intercept is also similar to 
the value ($\approx -3.6$) derived from 
\cite{trump15}, which is based on 
optically selected AGNs in the local 
universe ($z<0.1$).\footnote{\label{foot:jon} 
Starting from 
$\lambda_{\rm Edd}/\mathrm{sSFR} \approx 10^{-2.3}$\
in their Fig.~18 (where $\lambda_{\rm Edd}$\ is 
Eddington ratio and sSFR is specific SFR), 
we obtain $\rm BHAR/SFR \sim 10^{-3.6}$\ with
the assumption of $\epsilon \sim 0.1$\ 
(Sec.~\ref{sec:bhar}) and $M_*/M_\mathrm{BH} \sim 500$\
\citep[e.g.,][]{haring04, kormendy13}.}
In general, \xray\ emission from XRBs (the dashed lines) 
is lower compared to that from AGNs, and it 
is less significant at high SFR. 
The stacked \xray\ emission for individually 
undetected galaxies is consistent with 
being entirely due to XRBs.

\subsection{BHAR Dependence on $M_*$}
\label{sec:bhar_vs_m}
\begin{figure*}[htb]
\includegraphics[width=0.52\linewidth]{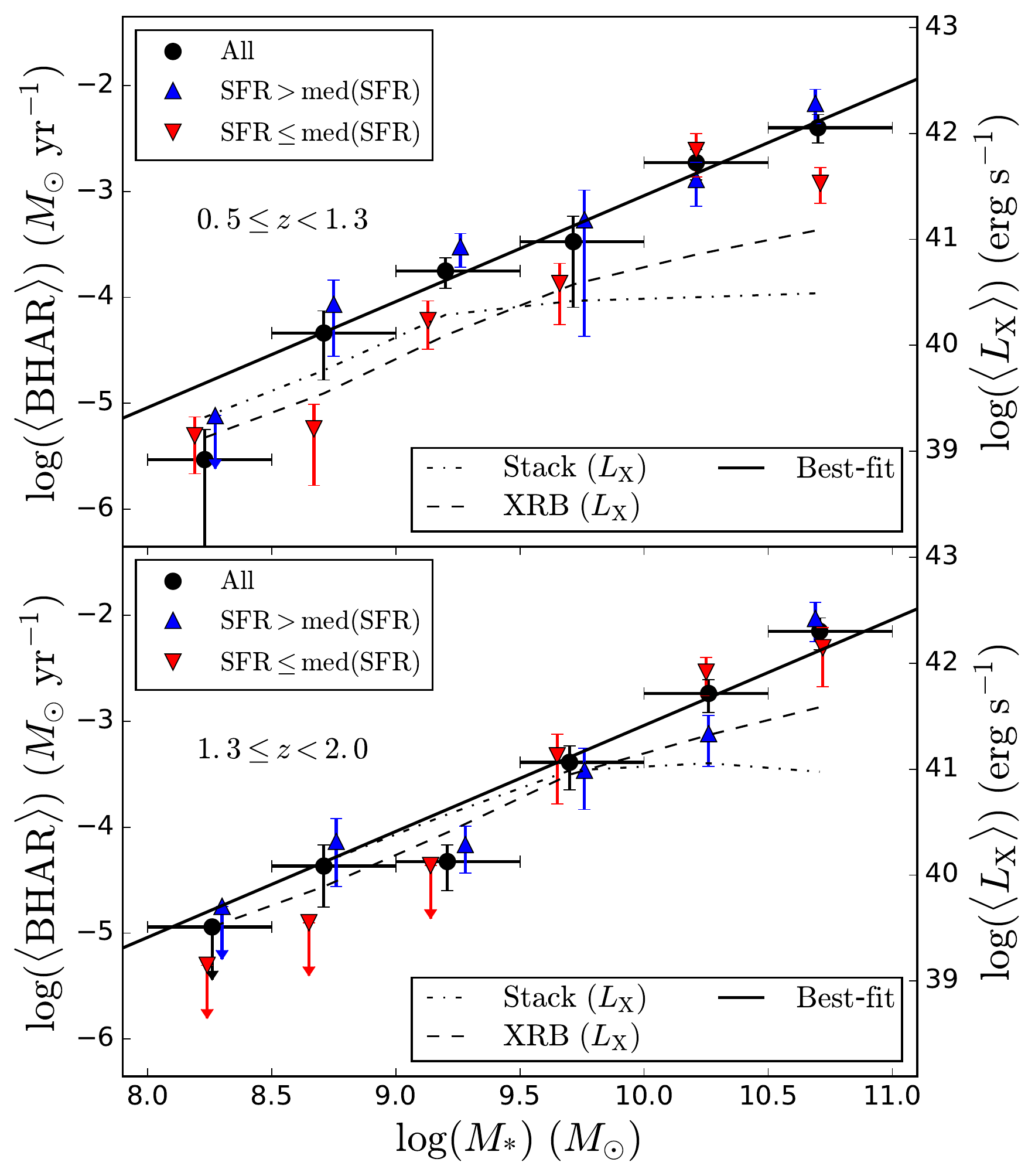}
\includegraphics[width=0.48\linewidth]{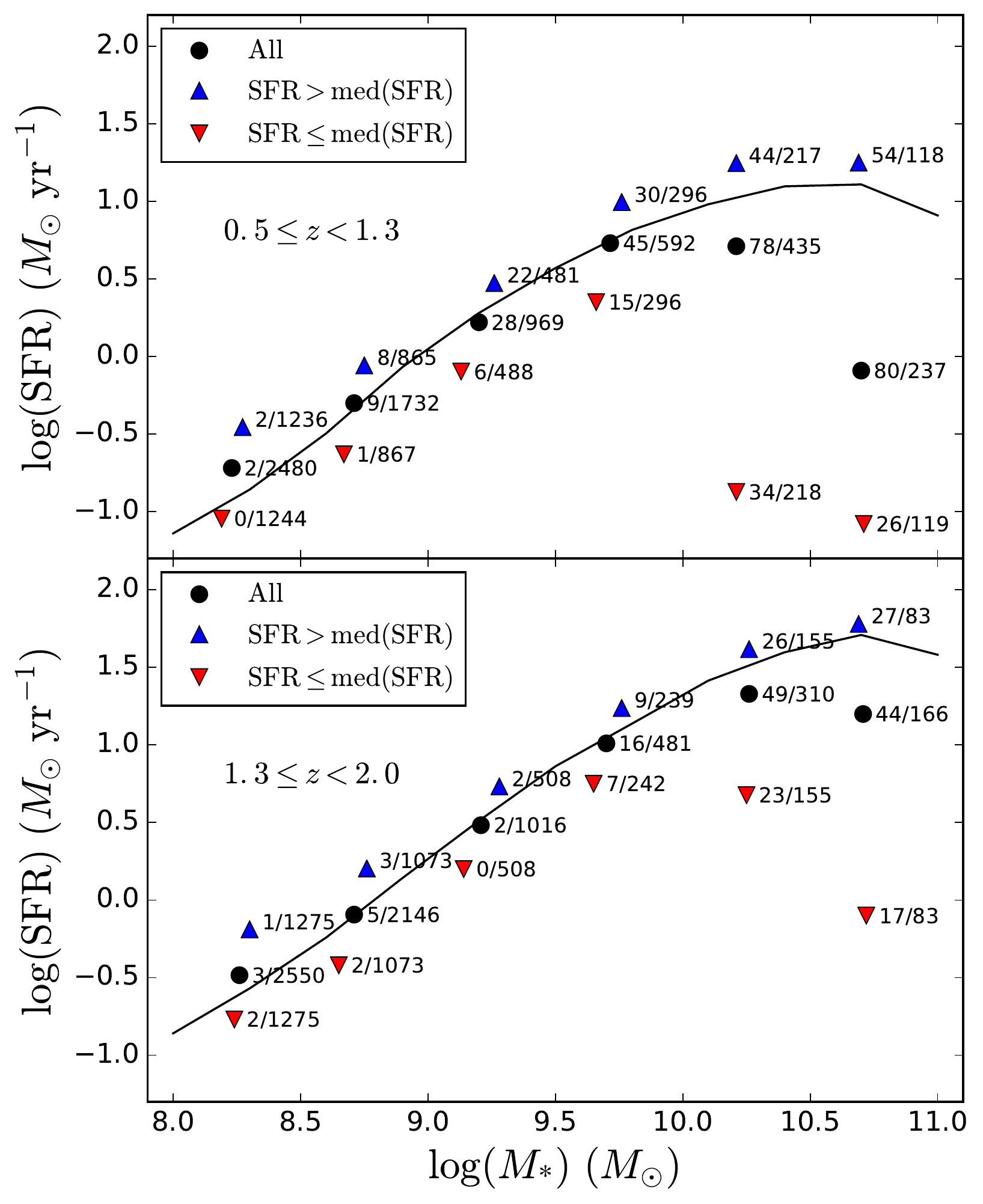}
\caption{Same format as Fig.~\ref{fig:Lx_vs_SFR}, but the 
bins are based on different $M_*$\ ($x$-axis), and 
each sample is split into two subsamples based 
on SFR. 
\bharm\ is positively correlated with $M_*$. 
The solid curves in the right panels indicate the
star-forming main sequence (B13).
}
\label{fig:Lx_vs_M}
\end{figure*}

To investigate the relation between \bharm\ and $M_*$, 
we bin our sources in $M_*$\ 
and calculate \bharm\ for each bin. The total number 
of sources is 13114 
(Tab.~\ref{tab:src_num}).\footnote{The sample size here 
is different from that in Sec.~\ref{sec:bhar_vs_sfr}, 
because the sample here is defined as 
$10^8 \leq M_* < 10^{11} M_\sun$\ while the sample in
Sec.~\ref{sec:bhar_vs_sfr} is defined as 
$0.1 \leq \mathrm{SFR} < 100 M_\sun$~yr$^{-1}$.}
The results are shown in Fig.~\ref{fig:Lx_vs_M}.
In general, \bharm\ is higher in more massive 
galaxies, and the fraction of \xray\ detected
sources rises toward higher $M_*$, consistent
with previous work (see Sec.~\ref{sec:intro}).
In the high-$z$\ bins, \xray\
emission from AGNs is comparable with that 
expected from XRBs for galaxies with 
$M_* < 10^{10}\ M_{\sun}$, but AGN emission 
becomes dominant for 
more massive galaxies. The fact that more massive 
galaxies have higher \bharm\ is also supported by  
Fig.~\ref{fig:Lx_vs_z} and Fig.~\ref{fig:m_vs_sfr}, 
which demonstrates that most \xray\ detections occur in 
galaxies with $M_* \gtrsim 10^{10}\ M_{\sun}$\ despite 
the fact that those massive galaxies are only 
$\approx 10\%$\ of the whole population.
Similar to the behavior in Fig.~\ref{fig:Lx_vs_SFR} 
left, \xray\ emission from stacked sources is 
generally comparable to that expected from XRBs.
There are several bins 
($M_* \lesssim 10^{9}\ M_{\sun}$) with only a few 
\xray-detected sources ($\lesssim 10$; see 
Fig.~\ref{fig:Lx_vs_M} right). 
In those bins, the \lxm\ contribution from
\xray-detected sources does not dominate 
over that from stacked sources; 
thus, the small numbers of detected sources 
do not cause large statistical fluctuations.

As in Sec.~\ref{sec:bhar_vs_sfr}, we perform
a linear fitting to the \bharm-$M_*$\ relation. 
If the slope is allowed to vary, we obtain
\begin{equation}\label{eq:bhar_m1}
\log (\langle \mathrm{BHAR} \rangle) 
=(1.10\pm0.08) \log (M_*) - (14.0\pm0.8)
\end{equation}
with reduced $\chi^2=0.88$\ (\hbox{$p$-$\mathrm{value}=54\%$});
if the slope is fixed to unity, we obtain
\begin{equation}\label{eq:bhar_m}
\log (\langle \mathrm{BHAR} \rangle) 
=\log (M_*) - (13.0\pm0.1)
\end{equation}
with reduced $\chi^2=0.94$\ (\hbox{$p$-$\mathrm{value}=50\%$}).
Therefore, as for the \bharm-SFR relation, the 
\bharm-$M_*$\ relation can also be described 
acceptably by a linear relation with 
a slope of unity. From Fig.~\ref{fig:Lx_vs_M} left, 
the \bharm-$M_*$\ relation is similar in both 
redshift ranges. 
The weak redshift dependence is 
consistent with the behaviors of the
AGN \xray\ luminosity function (XLF) and 
stellar mass function (SMF).
Both XLF and SMF of our studied regimes 
(i.e., $L_{\rm X} \lesssim 10^{44}$~erg~s$^{-1}$
and $10^8 \lesssim M_* \lesssim 10^{11} M_\sun$,
respectively)
drop slightly by $\approx 0.2$~dex 
from $z=0.8$\ to 1.7, 
where these redshift values are the medians
of the low-$z$\ and high-$z$\ samples, 
respectively 
(see, e.g., \citealt{ueda14, tomczak14}).

In the right panels of Fig.~\ref{fig:Lx_vs_M}, 
the median SFR values for 
the bins with $M_* \lesssim 10^{10}\ M_{\sun}$\ are 
close to those expected from the star-forming 
main sequence in the model of 
\citet[][B13 hereafter]{behroozi13}.
For massive galaxies with 
$M_* \gtrsim 10^{10}\ M_{\sun}$, our median SFRs 
are systematically lower than the  
values expected from the star-forming main sequence. 
This is likely due to the existence of massive 
evolved systems in our sample. 
In fact, after removing the quiescent population,
our median SFRs agree much better with the B13 model 
(see App.~\ref{app:sf_galaxy}).
For both the B13 model and our data, 
the SFR-$M_*$\
relation bends at $M_* \gtrsim 10^{10}\ M_\sun$\
(Fig.~\ref{fig:Lx_vs_M} right), 
likely due to the depletion of cold gas commonly 
found in high-$M_*$\ systems
\citep[e.g.,][]{peng15}.


\subsection{BHAR Dependence on Both SFR and $M_*$}
\label{sec:bhar_vs_sfr_and_m}
As shown in Secs.~\ref{sec:bhar_vs_sfr} and 
\ref{sec:bhar_vs_m}, \bharm\ has a positive dependence 
on both SFR and $M_*$. However, SFR and $M_*$\ are 
not independent properties for star-forming galaxies, 
which are the major population in our sample.
These two properties are positively 
related to each other via the star-formation main 
sequence (e.g., \citealt{elbaz11}; also see 
Fig~\ref{fig:m_vs_sfr} and the  
right panels of Figs.~\ref{fig:Lx_vs_SFR} and 
\ref{fig:Lx_vs_M}). 
Therefore, it is possible 
that \bharm\ is fundamentally correlated with 
one factor, and the observed relation with 
the other factor is only a secondary effect. 

To investigate this possibility, we compare
\bharm\ for sources with different $M_*$\ 
(SFR) but similar SFR ($M_*$). 
Specifically, we split each
sample in Sec.~\ref{sec:bhar_vs_sfr} into two 
subsamples, i.e., with $M_*\leq \med(M_*)$\ and 
$M_*> \med(M_*)$, respectively, where $\med(M_*)$\ 
is the median $M_*$\ in each sample.
For completeness, we also split the samples
with highest SFR, although the resulting 
subsamples have less than 100 sources
(but more than 50). 
The high-$M_*$\ subsamples have similar 
typical redshifts compared to the corresponding 
low-$M_*$\ subsamples; the differences between 
their median redshifts are $\approx 0.1$. 
We then calculate \bharm\ for both subsamples 
(red downward and blue upward triangles in 
Fig.~\ref{fig:Lx_vs_SFR}). In 
general, the high-$M_*$\ subsample has 
significantly higher ($\approx 0.5-1.5$~dex) 
\bharm\ than its low-$M_*$\ counterpart. 
The typical difference between 
the median $M_*$\ of 
the two subsamples is $\approx 0.5$~dex.

Similarly, we also divide each sample in each 
$M_*$\ bin (Sec.~\ref{sec:bhar_vs_m}) into two 
subsamples with 
$\mathrm{SFR} \leq \mathrm{med(SFR)}$\ 
and $\mathrm{SFR} > \mathrm{med(SFR)}$
(Fig.~\ref{fig:Lx_vs_M}). 
The high-SFR subsamples have slightly higher 
median redshifts than their low-SFR counterparts; 
the differences are $\approx 0.25$\ and $0.1$\ in 
the low-$z$\ and high-$z$\ ranges, respectively. 
This reflects the cosmic evolution of the 
star-forming main-sequence, i.e., galaxies tend to 
have higher specific SFR (sSFR, defined as SFR/$M_*$) 
at higher redshift. 

As shown in Fig.~\ref{fig:Lx_vs_M}, in the 
low-$z$\ range, the high-SFR subsample 
generally has higher \bharm, but for galaxies 
in the high-$z$\ range, the \bharm\ is  
similar for the two subsamples.  
Therefore, our results 
suggest that, at $1.3 \leq z < 2.0$, $M_*$\ 
is likely to be the main physical property   
correlated with \bharm, and the observed 
\bharm-SFR relation might be a secondary 
effect caused by the SFR-$M_*$\ correlation. 
This is also demonstrated by the comparison between
different (sub)samples in 
Fig.~\ref{fig:Lx_vs_M} bottom panels.
For example, the high-SFR subsample with 
$10^{9.5} < M_* < 10^{10}\ M_\sun$\ has median
SFR comparable to those of the two samples with 
highest $M_*$, but its \bharm\ value is much lower 
than those of the latter. 
Nevertheless, we cannot rule out 
possible correlation between \bharm\ and SFR
in the high-$z$\ range. 
This is because for bins with 
$M_* < 10^{10}\ M_{\sun}$, the \lxm\ from 
AGNs is low and comparable to 
\lxgm. Considering the uncertainties 
associated with, e.g., stacking procedures
and XRB modelling, our data are not sufficiently 
sensitive to differentiate possible \bharm\ differences 
between high-SFR and low-SFR samples in the 
low \bharm\ regime (when \xray\ emission 
from AGNs is comparable to or weaker than that 
from XRBs). 
It is thus possible that SFR 
correlates more strongly with \bharm\ at 
$M_* \gtrsim 10^{10.5}\ M_{\sun}$,
especially in the low-$z$\ range
(see the rightmost $M_*$\ bins in 
Fig.~\ref{fig:Lx_vs_M} left); the dependence 
of \bharm\ on SFR in the high-$M_*$\ regime
is also suggested by some previous studies  
\citep[e.g.,][]{delvecchio15, rodighiero15, trump15}.
\cite{rosario13} suggest that 
among massive galaxies with 
$M_* \gtrsim 10^{10.5}$~$M_\sun$, \xray\ AGNs
are more prevalent in high-SFR systems
(see also, e.g., \citealt{azadi15}). 
This is consistent with our observations. 
From the right panels of Fig.~\ref{fig:Lx_vs_M},
in the bins with $10^{10.5} \leq M_* < 10^{11}$~$M_\sun$, 
the high-SFR subsamples have $\approx 1.5-2$\ times
higher fractions of \xray-detected sources than
the low-SFR subsamples. 

\begin{figure}[htb]
\includegraphics[width=\linewidth]{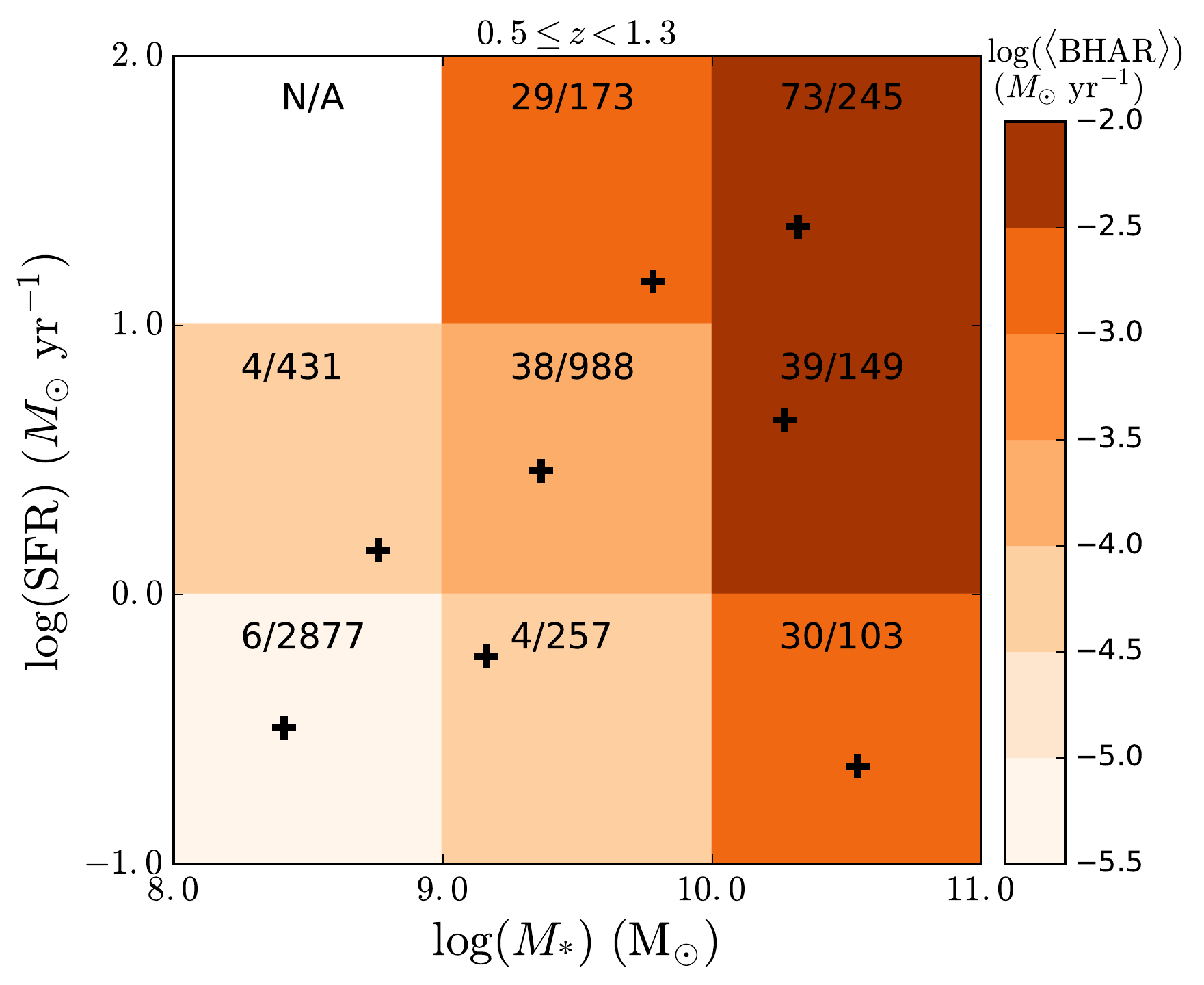}\\
\includegraphics[width=\linewidth]{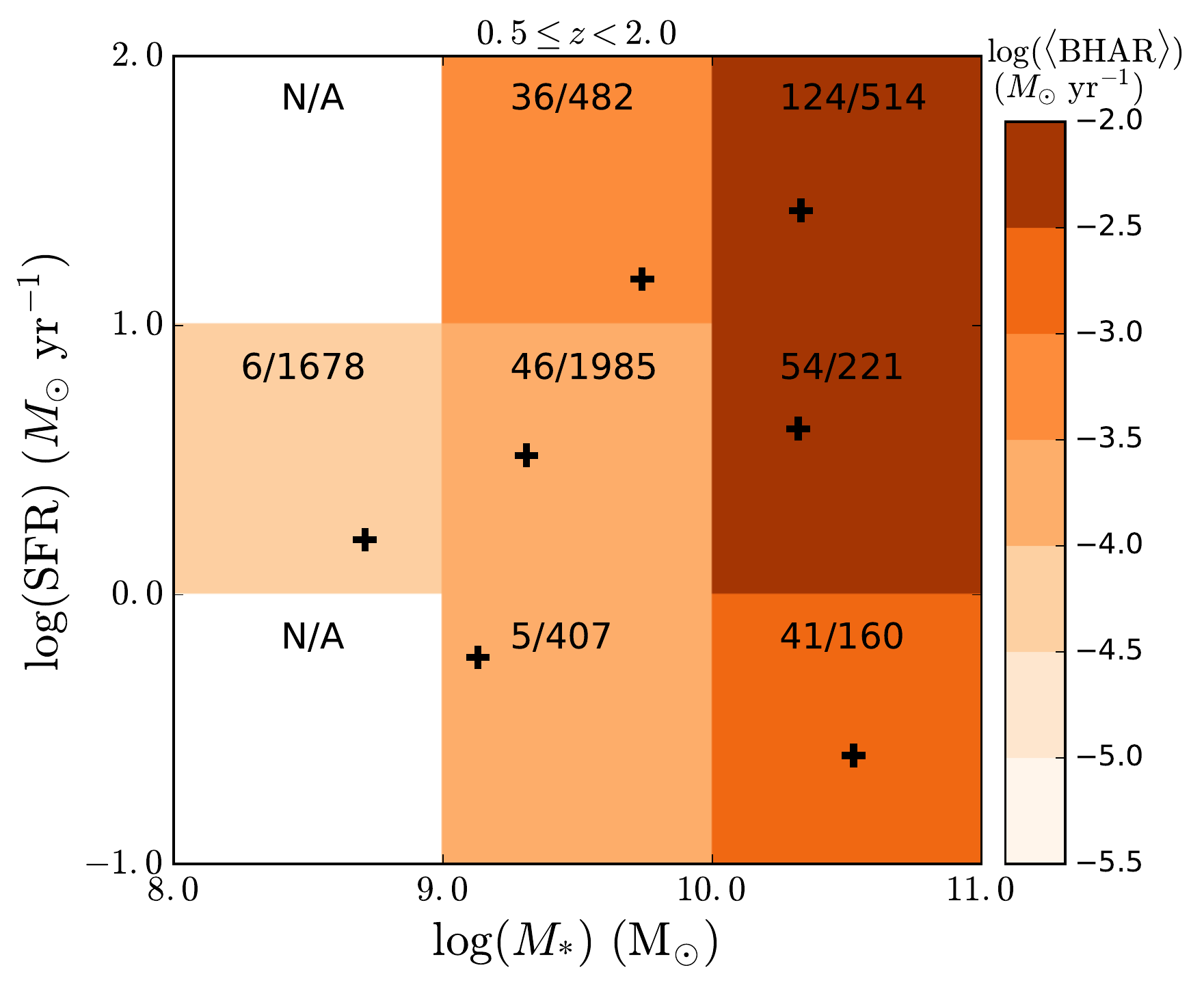}\\
\caption{Color-coded \bharm\ as a function of SFR and $M_*$\
at $0.5 \leq z < 1.3$\ (upper panel) and $0.5 \leq z < 2.0$\ 
(lower panel). 
The text in the center of each square indicates the number
of \xray\ detected sources and all sources. 
The black cross indicates median values of SFR and $M_*$\ 
for sources in each bin, which are adopted in the analyses
in Sec.~\ref{sec:bhar_vs_sfr_and_m}. 
In both panels, the upper-left
squares with white color include too few sources 
(fewer than 100; see Sec.~\ref{sec:bhar_vs_sfr}),
and their \bharm\ values are not calculated.
In the lower panel, the \bharm\ of the bottom-left
square has a negative $1\sigma$\ lower limit; its 
value is not shown.
The results for the two redshift bins are 
similar. Both $M_*$\ and SFR
have positive correlations with \bharm. 
}
\label{fig:BHAR_vs_M_and_SFR}
\end{figure}

To clarify better whether SFR or $M_*$\  
is more important in the low-$z$\ range, we 
bin our sources with $0.5 \leq z < 1.3$\ 
over grids of SFR 
($0.1 \leq \mathrm{SFR} < 100\ 
M_{\sun}\ \mathrm{yr}^{-1}$) and $M_*$\ 
($10^{8} \leq M_* < 10^{11}\ M_{\sun}$) 
with the number of sources totaling 5224 
(see Tab.~\ref{tab:src_num}).
We enlarge the bin width to 1~dex to 
include more sources in each bin and reduce the
uncertainties of \bharm\ measurements
(see Fig.~\ref{fig:m_vs_sfr}). 
The larger bin width also makes our results 
less sensitive to the measurement errors
on SFR and $M_*$. 
The results are shown in the upper panel of 
Fig.~\ref{fig:BHAR_vs_M_and_SFR}. 
As expected, both SFR and $M_*$\ display 
positive correlations with \bharm. 

We perform partial-correlation 
(PCOR) analyses on the results 
using {\sc ``pcor''} in R.\footnote{The
R code {\sc pcor} is available from 
http://www.yilab.gatech.edu/pcor.html
\citep[e.g.,][]{johnson02}. 
In the analyses, we provide {\sc pcor} with 
$\log(\langle \rm BHAR \rangle)$, $\log(\rm SFR)$, 
and $\log(M_*)$\ for sources in each bin 
(as indicated by the black crosses in 
Fig.~\ref{fig:BHAR_vs_M_and_SFR}). 
Here, we use a logarithmic scale instead of 
a linear scale to reduce potential power-law 
relations to linear relations among 
the three quantities. 
}
The PCOR analyses are deployed  
to measure the correlation between \bharm\ 
and SFR ($M_*$) while controlling for the 
effects of $M_*$\ (SFR). 
There are three statistics available in pcor 
to perform the analyses: one parametric
statistic (Pearson) and two non-parametric 
statistics (Spearman and Kendall). 
We perform the analyses with all three methods
and list the $p$-values from each method in
Tab.~\ref{tab:pca}. All $p$-values 
for the \bharm-$M_*$\ relation are significantly 
smaller than the corresponding $p$-values 
for the \bharm-SFR relation, indicating that \bharm\ 
correlates with $M_*$\ more strongly than
SFR. The parametric method produces $p$-values 
generally smaller than the non-parametric 
methods. This is because the parametric method
assumes linear relations (see 
Eqs.~\ref{eq:bhar_sfr1}, \ref{eq:bhar_sfr}, 
\ref{eq:bhar_m1}, and \ref{eq:bhar_m}; see also 
Eqs.~\ref{eq:bhar_sfr_m1} and \ref{eq:bhar_sfr_m2}
below) and uses the input data quantitatively, 
while the non-parametric methods do not have such 
assumptions but only use ranks of the input data. 
Among the non-parametric methods, Spearman's 
statistic leads to more significant relations 
than Kendall's statistic. The reason is likely
that the former uses the value of the 
difference between two ranks, 
while the latter is even more conservative, and 
only considers the sign of the difference 
(e.g., \citealt{feigelson12}).
A linear regression results in
\begin{equation}\label{eq:bhar_sfr_m1}
\begin{split}
\log (\langle \mathrm{BHAR} & \rangle) 
=(0.26\pm0.11) \log (\mathrm{SFR}) \\
&+(1.24\pm0.11) \log (M_*)-(15.5\pm 1.1)
\end{split}
\end{equation}
with reduced $\chi^2=0.51$\ (model-rejection 
\hbox{$p$-$\mathrm{value}=77\%$};
see Footnote~\ref{foot:fit} for 
the fitting method).

\begin{figure}[htb]
\includegraphics[width=\linewidth]{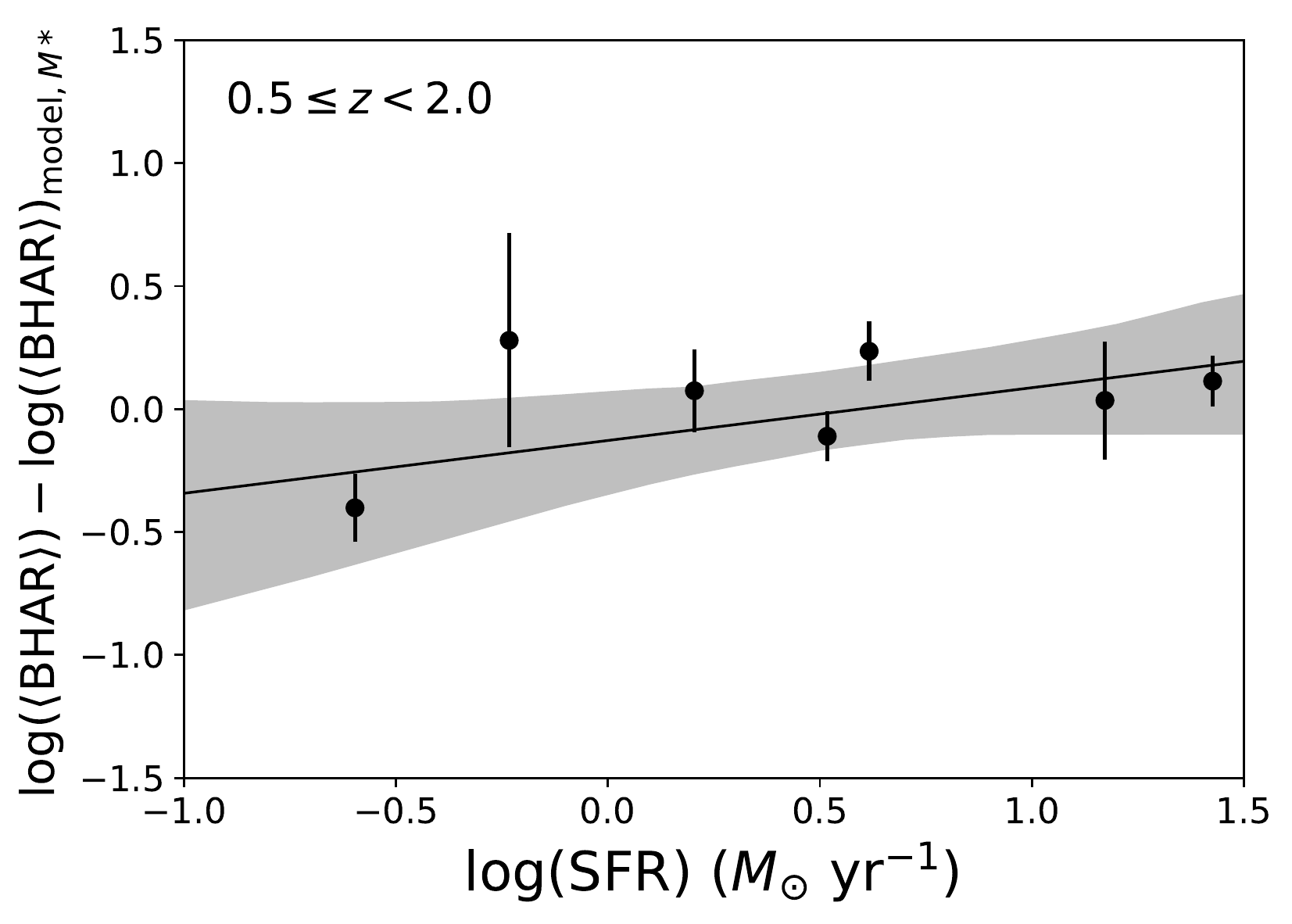}\\
\includegraphics[width=\linewidth]{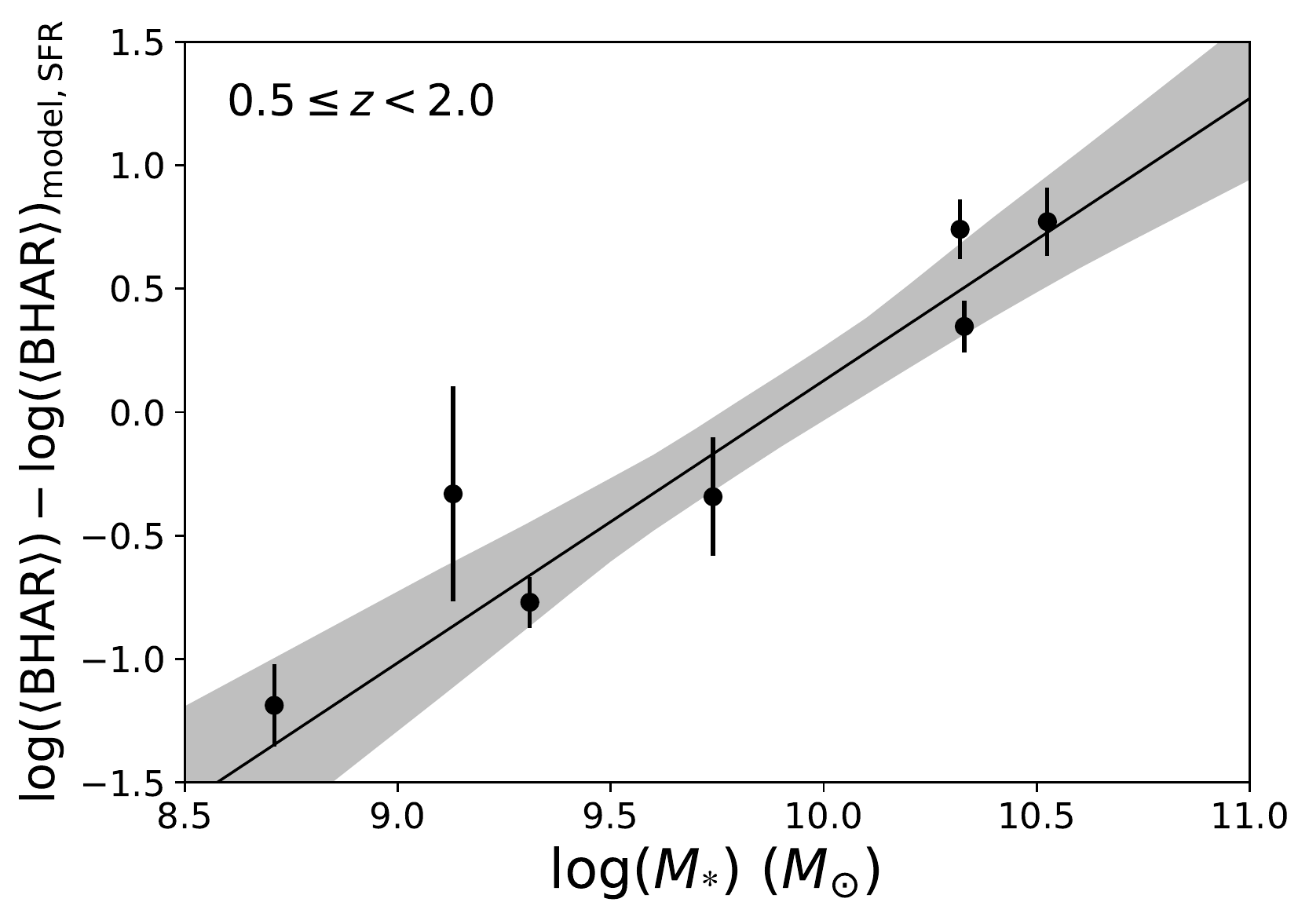}
\caption{Top: the 
$\langle \mathrm{BHAR} \rangle$ 
residuals of the 
$\langle \mathrm{BHAR} \rangle$-$M_*$\ 
fit as a function of SFR. The black solid line indicates 
the best fit for the residuals as a function of SFR. 
The shaded region indicates $3\sigma$\ 
uncertainties derived from Markov chain Monte Carlo 
sampling utilizing {\sc ``emcee''} \citep{mackey13}. 
The residual-SFR relation is flat.
Bottom: the $\langle \mathrm{BHAR} \rangle$ 
residual of the $\langle \mathrm{BHAR} \rangle$-SFR\ 
fit as a function of $M_*$.
The residual-$M_*$\ relation is steep. 
}
\label{fig:residual}
\end{figure}

We also perform the same analyses for 
sources in both the low-$z$\ and 
high-$z$\ ranges together 
(11460 sources with $0.5\leq z < 2.0$\ as listed in 
Tab.~\ref{tab:src_num}).\footnote{
We do not analyze the high-$z$\ range 
independently, because this would lead 
to only three available bins 
having $>100$~sources 
(see Sec.~\ref{sec:bhar_vs_sfr}) 
and well-constrained \lx\ (positive $1\sigma$\ 
lower limit); such a small number of bins 
is not suitable for PCOR analyses.
}
Fig.~\ref{fig:BHAR_vs_M_and_SFR} (bottom) 
displays the results. 
The $p$-values from PCOR analyses 
are shown in Tab.~\ref{tab:pca}. 
The \bharm-$M_*$\ relation is still 
significant, in qualitative agreement with 
the results for the low-$z$\ range, but the 
\bharm-SFR relation becomes insignificant.
To visualize the PCOR analyses, we first 
fit $\langle \mathrm{BHAR} \rangle$\ 
as a linear function of $M_*$\ (SFR) 
in logarithmic space, using the same data as
in the PCOR analyses (see 
Fig.~\ref{fig:BHAR_vs_M_and_SFR} bottom).
Then we model the residuals as a linear function 
of SFR ($M_*$), and show the results 
in Fig.~\ref{fig:residual}. 
The resulting residual-SFR relation is 
flat, and its slope is consistent with zero
at a $3\sigma$\ confidence level.
However, the residual-$M_*$\ 
relation is steep.
Therefore, $\langle \mathrm{BHAR} \rangle$\ 
can largely be described via 
the relation with $M_*$\ rather than SFR. 
Similar analyses have also been performed for the 
low-redshift bin, and the conclusion is the 
same.

The linear-fitting 
($\langle \mathrm{BHAR} \rangle$\ 
as a function of SFR and $M_*$) result is  
\begin{equation}\label{eq:bhar_sfr_m2}
\begin{split}
\log (\langle \mathrm{BHAR} & \rangle) 
 = (0.22\pm0.08) \log (\mathrm{SFR}) \\
 & +(1.16\pm0.09) \log (M_*)-(14.6\pm 0.8),
\end{split}
\end{equation}
similar as in the low-$z$\ range.
The best-fit reduced $\chi^2$\ is 2.0
which corresponds to a
\hbox{$p$-$\mathrm{value}=9\%$}. 
The \hbox{$p$-value} is much smaller 
than the previous value because the 
errors on \bharm\ from $0.5 \leq z < 2.0$\
($0.10-0.44$~dex with $\rm median=0.14$~dex)
are generally smaller than those 
from $0.5 \leq z < 1.3$\ 
($0.15-0.48$~dex with $\rm median=0.22$~dex) 
due to the increase of sample size in each bin 
(Fig.~\ref{fig:BHAR_vs_M_and_SFR}).

The above analyses are based on 
samples including both star-forming and 
quiescent galaxies, with star-forming galaxies
being the major population ($\approx 80\%$).
To test if our main conclusion (\bharm\ 
mainly relates to $M_*$) applies for 
star-forming galaxies alone, we repeat
the above analyses with the sample of galaxies
near the star-forming main sequence in 
App.~\ref{app:sf_galaxy}. Our analyses there show
\bharm\ still correlates with $M_*$\ more 
strongly than SFR for star-forming galaxies. 

\begin{table}
\begin{center}
\caption{$p$-values (Significances) of 
	Partial Correlation Analyses}
\label{tab:pca}
\begin{tabular}{cccc}\hline\hline
\multicolumn{4}{c}{$0.5 \leq z < 1.3$} \\ \hline
Relation & Pearson & Spearman & Kendall \\
BHAR-SFR &  $10^{-6}\ (4.8\sigma)$ &  
	    0.01$\ (2.6\sigma)$  & 
	    $0.34\ (0.9\sigma)$ \\
BHAR-$M_*$ &  $10^{-105}\rm \ (22\sigma)$ & 
	      $10^{-12} \ (7.1\sigma)$   & 
	      $0.01\ (2.6\sigma)$ \\ \hline\hline
\multicolumn{4}{c}{$0.5 \leq z < 2.0$} \\ \hline
Relation & Pearson & Spearman & Kendall \\
BHAR-SFR &  $0.30\ (1.0\sigma)$ &  
	    $0.23\ (1.2\sigma)$ & 
	    $0.68 \ (0.4\sigma)$  \\
BHAR-$M_*$ & $10^{-14} \ (7.5\sigma)$ & 
	     $10^{-4} \ (3.5\sigma)$ & 
	     $0.06\ (1.9\sigma)$ \\ \hline
\end{tabular}
\end{center}
\end{table}

\subsection{BHAR/SFR Ratio as a Function of $M_*$}
\label{sec:bhar_ov_sfr_vs_m} 
\begin{figure}[htb]
\includegraphics[width=\linewidth]{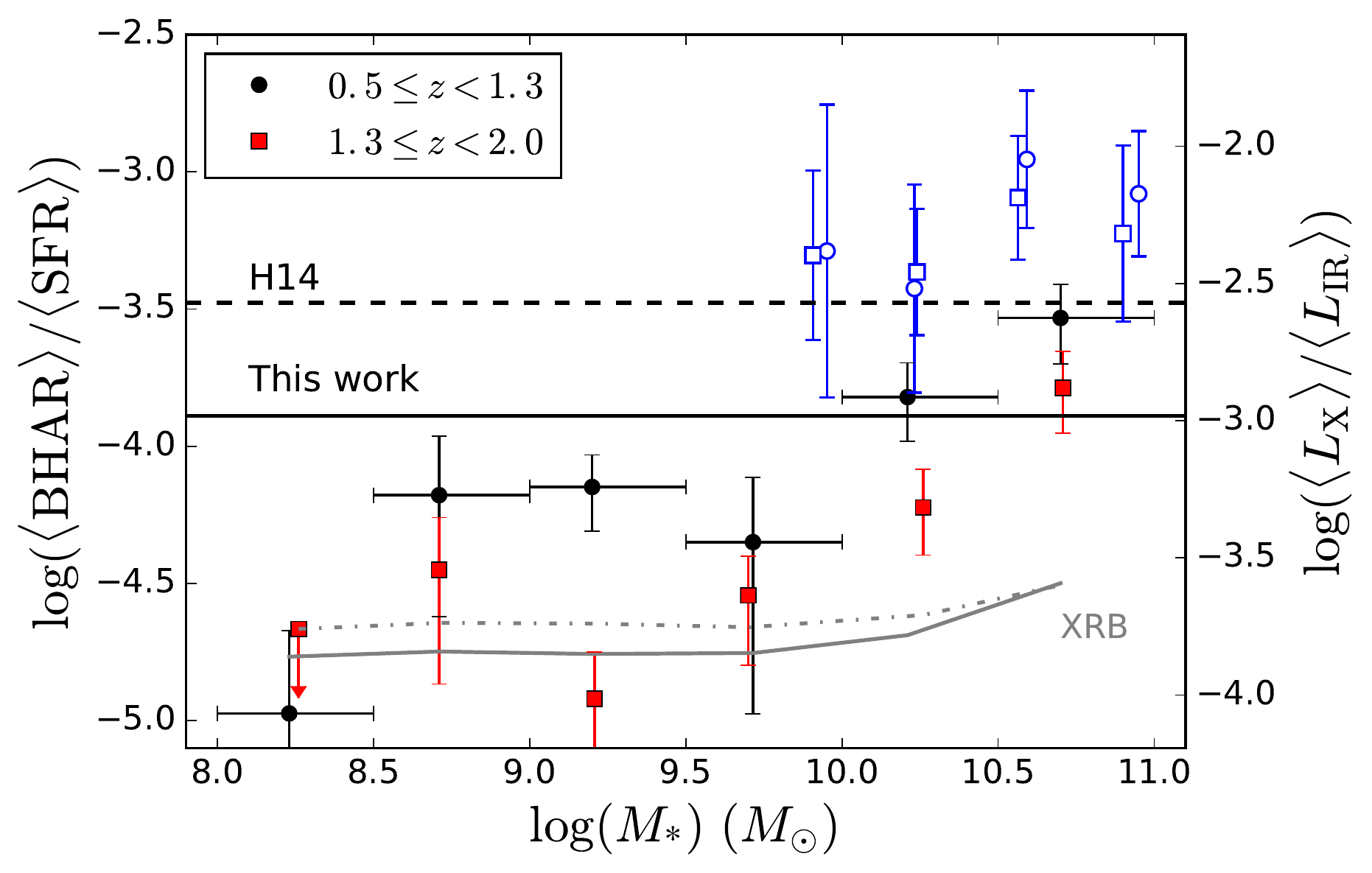}\\
\caption{The ratio of \bharm\ to \sfrm\ as a function 
of $M_*$. The black points and red squares indicate sources
in the low-$z$\ and high-$z$\ ranges, 
respectively. Their horizontal 
positions indicate median $M_*$\ in the bin, and the 
horizontal error bars (only shown for the low-$z$\ bin) 
indicate the bin width. The blue open symbols 
indicate results from \cite{mullaney12}; the circles 
and squares indicate their $z\sim 1$\ and $z \sim 2$\
samples, respectively. The scale 
of \hbox{\lxm$/\langle L_{\rm IR} \rangle$} 
is shown on the right 
side, where we convert SFR\ to $L_{\rm IR}$\ with
a constant factor (C13).
The gray solid and dash-dotted lines indicate the 
\hbox{\lxgm$ / \langle L_{\rm IR} \rangle$} 
ratios in the low-$z$\ and high-$z$\ ranges,
respectively. 
These values are used as upper limits for 
\hbox{\lxm$ / \langle L_{\rm IR} \rangle$} for AGNs, 
when 1$\sigma$\ lower 
limits of \lxm\ for AGNs are negative.
The solid horizontal line indicates the intercept 
of our best-fit \bharm-SFR linear relation in
Fig.~\ref{fig:Lx_vs_SFR}; the dashed horizontal 
line indicates the H14 model.
The \bharm/\sfrm\ ratio is strongly dependent on $M_*$. 
More massive galaxies generally have higher 
\bharm/\sfrm, indicating that they are more 
efficient in growing their SMBHs.
For high-mass galaxies 
($M_* \gtrsim 10^{10}\ M_\sun$), our 
$\langle \mathrm{BHAR} \rangle$/$\langle \mathrm{SFR} \rangle$\ 
values are systematically lower than those in \cite{mullaney12}.
This is likely caused by the exclusion of broad-line 
AGNs in our sample (see Secs.~\ref{sec:samp} and 
\ref{sec:missed_bhar}).
}
\label{fig:ratio_vs_M}
\end{figure}

The ratio BHAR/SFR represents the relative 
growth between SMBHs and their host galaxies,
and thus it has important implications 
for SMBH-galaxy coevolution. 
To study its dependence 
on $M_*$, we bin our sources based on 
$M_*$\ and derive \bharm/\sfrm\ for each bin.
The results are displayed in 
Fig.~\ref{fig:ratio_vs_M}.
For both redshift ranges, massive galaxies with
$M_* \gtrsim 10^{10}\ M_\sun$\
generally have higher \bharm/\sfrm. 
This is understandable considering that \bharm/$M_*$\
is roughly a constant (Sec.~\ref{sec:bhar_vs_m}), 
while sSFR generally drops for massive galaxies
(Fig.~\ref{fig:Lx_vs_M} right; also see, e.g., 
\citealt{whitaker12, pan16}).

However, the \bharm/\sfrm\ dependence on $M_*$\ is 
not observed by \citet[][see the blue open 
points in Fig.~\ref{fig:ratio_vs_M}]{mullaney12}.
This difference is likely because their sample consists 
of only massive galaxies in a narrow range of $M_*$ 
($10^{10} \lesssim M_* \lesssim 10^{11}\ M_{\sun}$)
and their uncertainties are relatively large. A 
recent study by \citet{rodighiero15}, based on galaxies
with $10^{10} \lesssim M_* \lesssim 10^{11.5}\ M_{\sun}$, 
found that a positive 
correlation still exists between \bharm/\sfrm\ and $M_*$, 
due to their small error bars on \bharm/\sfrm. 
Our \bharm/\sfrm\ values for galaxies having similar $M_*$\
are slightly lower than those measured in the two studies,
likely due to different sample selections and/or 
the missed accretion power from broad-line AGNs 
in our analyses 
(Secs.~\ref{sec:samp} and \ref{sec:missed_bhar}). 
Indeed, if we assume the luminous broad-line AGNs
have $10^{10} \leq M_* < 10^{11}\ M_{\sun}$\ 
and include them in our sample, our 
$\langle \mathrm{BHAR} 
\rangle$/$\langle \mathrm{SFR} \rangle$\ 
values would be consistent with those in previous studies 
(see Sec.~\ref{sec:missed_bhar} for details).
The $M_*$-dependent ratio of \bharm/\sfrm\ provides a possible 
explanation for the fact that our best-fit intercept 
of the \bharm-SFR relation is lower than that expected 
from the H14 model (see Sec.~\ref{sec:bhar_vs_sfr}). 
The data used by H14 to estimate the \bharm-SFR relation
are mainly for massive galaxies. Since massive 
galaxies have higher \bharm/\sfrm\ values, the resulting 
intercept of the \bharm-SFR relation should be higher.
For example, if we only use the high-$M_*$\ subsamples 
(blue points in Fig.~\ref{fig:Lx_vs_SFR}) to derive
the \bharm-SFR relation with unity slope, we 
would obtain a higher intercept. 

The \bharm/\sfrm\ ratio in the high-$z$\ range is
generally lower than that in the low-$z$\ range 
(Fig.~\ref{fig:ratio_vs_M}). This is consistent with
global AGN activity and star formation studies. 
The emissivity of AGNs in our luminosity regime 
($L_{\rm X} \lesssim 10^{44}$~erg~s$^{-1}$; see 
Fig.~\ref{fig:Lx_vs_z}) slightly increases from 
$z \approx 2$\ to $z \approx 1$\ 
\citep[e.g.,][]{ueda14}; meanwhile, the 
emissivity of star formation drops 
(e.g., \citealt{hopkins04} and B13; see also 
Fig.~\ref{fig:Lx_vs_M} right).
Physically, the redshift evolution might reflect 
that, at lower redshift, gas in galaxies is more 
concentrated in the vicinity of SMBHs.



\subsection{Reliability Checks}
\subsubsection{Missed Accretion Power}
\label{sec:missed_bhar}
Luminous \xray\ emission is almost a universal 
tracer of SMBH accretion 
\citep[e.g.,][]{mushotzky04, gibson08}. 
Since this work is based on the deepest \xray\ survey
(the 7~Ms \cdfs) and includes individually 
undetected sources via stacking analyses 
(Sec.~\ref{sec:bhar}), we are unlikely 
to miss a large fraction of black-hole accretion power
due to survey sensitivity \citep[e.g.,][]{brandt15}. 
Also, our stacked mean \xray\ luminosities 
are similar to the predicted \xray\ emission 
from XRBs (see Secs.~\ref{sec:bhar_vs_sfr} 
and \ref{sec:bhar_vs_m}), 
indicating that most of cosmic accretion power 
in our redshift range ($0.5 \leq z < 2.0$)
is directly detected in the 7~Ms \cdfs. 
Due to the small size of \goodss\ 
(170~arcmin$^2$), our sample will 
miss AGNs at the bright end of the XLF
($L_{\rm X} \gtrsim 10^{44}$~erg~s$^{-1}$;
see Fig.~\ref{fig:Lx_vs_z}). 
For the \goodss\ field ($170\ \rm arcmin^2$), 
the fraction of missed accretion power is 
$\approx 37\%$\ for $0.5 \leq z < 2.0$,
estimated based on the XLF model of \cite{aird10}.
However, those 
luminous sources are likely to reside in massive
systems ($M_* \gtrsim 10^{10}\ M_\sun$; 
e.g., \citealt{bahcall97, kauffmann03, matsuoka14}), 
and thus the bright-end correction 
is not likely to boost \bharm\ for low-$M_*$\ 
galaxies and affect our main conclusions
(i.e., massive galaxies have higher \bharm\
and \bharm/\sfrm\ than less-massive ones).
Moreover, a recent study of luminous quasars 
also suggests their BHAR might be primarily related to 
$M_*$\ rather than SFR \citep[][]{xu15}.

A large population of CTK AGNs 
might exist but show almost no \xray\ signal, 
even in the unprecedentedly 
deep 7~Ms \cdfs\ (e.g., see Sec.~3.3 of 
\citealt{brandt15}; \citealt{comastri15}). 
Their accretion power 
would be largely missed in our analyses.
We have stacked hard-band 
(observed-frame $2-7$~keV) \xray\ images
and compared the results with the stacking of 
the soft band (observed-frame $0.5-2$~keV). 
We do not find evidence of 
CTK populations. Nevertheless, it is possible
to find signatures of CTK AGNs with more 
refined analyses (e.g., stacking of very hard
narrower bands), and we will perform such 
analyses in a dedicated paper (B.~Luo et al., 
in preparation).
The prevalence of CTK AGNs is 
predicted by population-synthesis models of 
the cosmic \xray\ background, although
these models have significant uncertainties 
in the regime of very high \nh\ 
\citep[e.g.,][]{gilli07, ueda14}. 
The predicted total \xray\ emission from CTK 
AGNs is usually less than that from other 
AGNs (e.g., \citealt{gilli07, treister09}; also
see \citealt{buchner15} for relevant spectral 
fitting), consistent with the 
very-hard \xray\ ($\gtrsim 10$~keV) observations 
of the local universe \citep[e.g.,][]{akylas16, koss16}.
Even if these CTK AGNs predominantly 
reside in low-$M_*$\ systems, the total \lx\
from these galaxies will be at most the same
as that from high-$M_*$\ galaxies; the 
average \lx\ in low-$M_*$\ galaxies should be 
still much lower than for massive galaxies, 
as the former has much larger number density.
Their inclusion is unlikely to make qualitative 
changes to our main conclusion that \bharm\ 
is a strong function of $M_*$.
Nevertheless, the possible existence of the
CTK AGN population could increase/decrease
the dependence on other galaxy properties
at given $M_*$.
Some observations suggest that CTK AGNs are
more prevalent in galaxies with high SFRs
and/or that are experiencing major mergers 
\citep[e.g.,][]{juneau13, kocevski15}, 
while the dependence on redshift tends 
not to be strong \citep[e.g.,][]{buchner15}.

Nineteen broad-line AGNs have been 
deliberately excluded from 
our sample (see Sec.~\ref{sec:samp}). The 
majority of them (14/19) are \xray\ luminous 
with $10^{43} < L_{\rm X} \lesssim 10^{44}$~erg~s$^{-1}$\
(see Fig.~\ref{fig:Lx_vs_z} bottom).
The rarity of broad-line AGNs with 
$L_{\rm X} < 10^{43}$~erg~s$^{-1}$\ is consistent 
with, e.g., \cite{merloni14}.  
The 14 luminous broad-line AGNs 
make up of 25\% of the total AGN population 
with $L_{\rm X} > 10^{43}$~erg~s$^{-1}$.
This fraction agrees with Fig.~7 of 
\cite{merloni14}, which shows the fraction of 
broad-line AGNs is $\approx 20-40\%$\ for 
$10^{43} \lesssim L_{\rm X} 
\lesssim 10^{44}$~erg~s$^{-1}$. Therefore, 
we are not missing a significant fraction 
of broad-line AGNs due to, e.g., low-S/N 
spectra.  
Those luminous broad-line AGNs are likely 
to reside in massive galaxies 
\citep[e.g.,][]{bahcall97, matsuoka14}. 
Indeed, $\approx 90\%$\ of our non-broad-line 
AGNs in the same luminosity regime are hosted 
by galaxies with $M_*>10^{10}\ M_{\sun}$,
and these should be physically similar 
systems as broad-line AGNs following 
the standard unified AGN model.
Hence, including the broad-line AGNs
would boost \bharm\ for massive galaxies,
and thus would make our main conclusions 
even stronger. 
This might also explain why our \bharm/\sfrm\ 
ratio for massive galaxies is lower than the 
values from previous studies 
(see Secs.~\ref{sec:bhar_vs_sfr} and 
\ref{sec:bhar_ov_sfr_vs_m}).
Quantitatively, if we include
those 14 luminous broad-line AGNs assuming 
they reside in galaxies with 
$10^{10} \leq M_* < 10^{11}\ M_\sun$,
the $\langle \mathrm{BHAR} \rangle / 
\langle \mathrm{SFR} \rangle$\ values 
for this $M_*$\ bin
would be $10^{-3.3}$ and $10^{-3.7}$, 
respectively, in the low-$z$\ and 
high-$z$\ ranges.
These values are consistent with 
previous studies (see Fig.~\ref{fig:ratio_vs_M}).
Considering the small population and relatively
low luminosities ($L_{\rm X}<10^{43}$~erg~s$^{-1}$) 
of the remaining 5 broad-line AGNs, excluding 
them is not likely to affect our results 
significantly.

\subsubsection{Sample $M_*$\ and SFR Completeness}
\label{sec:complete}
In the high-$z$\ range, our sample is not 
complete at $M_* \sim 10^{7}\ M_\sun$ 
(Fig.~\ref{fig:Lx_vs_z} top), and this is the 
main reason why this study focuses on 
galaxies with $M_* \gtrsim 10^{8}\ M_\sun$. 
To check quantitatively if our sample is complete 
down to $M_* \sim 10^{8}\ M_\sun$ in the high-$z$\ 
range, we calculate the comoving number density 
for the bin with lowest $M_*$\ and compare it 
with the model of B13. 
There are 2550 sources in 
the bin (see Fig.~\ref{fig:Lx_vs_M}); 
the comoving volume covered by the 
\goodss\ field (170~arcmin$^2$; see 
\citealt{guo13}) is 
$3.8 \times 10^{5}$~Mpc$^3$. Thus, 
the comoving number density for our 
galaxies having 
$10^8 \leq M_* < 10^{8.5}\ M_{\sun}$\ is 
$1.3 \times 10^{-2}$~Mpc$^{-3}$~dex$^{-1}$.
This value is roughly consistent with Fig.~3 of 
B13, indicating our sample is basically complete 
above $M_* \sim 10^{8}\ M_{\sun}$. 

A color-dependent completeness 
issue might exist in our flux-limited sample 
\citep[e.g.,][]{xue10}. This is because 
for a given $M_*$, blue galaxies generally 
have higher optical-to-near-IR 
luminosities than red galaxies
due to their relatively young stellar 
populations. Therefore, flux-limited 
optical/IR surveys like CANDELS might 
miss the red population in the low-$M_*$\
regimes. However, our sample is not likely
to have this issue, since our SFR-$M_*$\ relation 
agrees well with the B13 model
at the low-$M_*$\ end (see the right 
panels of Fig.~\ref{fig:Lx_vs_M}). 
If our sample were biased to blue 
(high-SFR) galaxies at the low-$M_*$\ 
end, the measured SFR would be 
significantly above the model value. 


The fact that our sample generally does 
not have color-dependent completeness 
is not surprising.
At the low $M_*$\ ($\sim 10^8\ M_{\sun}$) 
regime that we probe, the spread in 
rest-frame colors is relatively small,
i.e., most low-$M_*$\ (thus low-luminosity)
galaxies reside in the ``blue cloud''
rather than the ``red sequence'' in
color-magnitude diagrams 
\citep[e.g.,][]{schneider14, martis16, pan16}. 
This is broadly consistent with the 
right panels of Fig.~\ref{fig:Lx_vs_M},
assuming star-forming activity 
is generally traced by colors.  
The SFR dispersion is only 
$\lesssim 0.3$~dex in the bins whose 
$M_* \lesssim 10^{10}\ M_{\sun}$,
i.e., most galaxies are located around the 
star-forming main sequence. 
In the high-$M_*$\ regime where the 
red sequence exists, our sample is also likely
to be complete. This is because the comoving
number density is 
$8.5\times 10^{-4}$~Mpc$^{-3}$~dex$^{-1}$\ 
for the $10^{10.5} \leq M_* < 10^{11}\ M_{\sun}$\ 
bin in the high-$z$\ range; this value is
consistent with the stellar-mass function 
derived in a dedicated study
($9.3^{+1.9}_{-1.7}\times 10^{-4}$~Mpc$^{-3}$~dex$^{-1}$; 
see Tab.~1 of \citealt{tomczak14}), which 
includes both star-forming and quiescent 
galaxies. 

On the other hand, our lowest-SFR regime 
($\mathrm{SFR}\ \sim 0.1\ 
M_\sun$~yr$^{-1}$) corresponds to 
$M_* \sim 10^{8}\ M_\sun$\ on
the star-forming main sequence
(see the right panels of 
Figs.~\ref{fig:Lx_vs_SFR} and \ref{fig:Lx_vs_M}). 
Since our sample is basically complete 
above $M_* \sim 10^{8}\ M_\sun$, 
it should also be roughly complete 
above $\mathrm{SFR} \sim 0.1\ M_\sun$~yr$^{-1}$.
Moreover, our conclusions do not critically 
depend on the galaxies with
$\mathrm{SFR} \sim 0.1\ M_\sun$~yr$^{-1}$\ 
and $M_* \sim 10^{8}\ M_\sun$. Thus, 
even if there are minor completeness issues
in our lowest-SFR and/or lowest-$M_*$\ bins,
our results should not be affected 
materially. 


\section{Discussion}\label{sec:disc}
\subsection{SMBH-Galaxy Coevolution}
\label{dis:coev}
The linear relation between \bharm\ and SFR was 
previously derived for observations of high-SFR galaxies 
($\mathrm{SFR} \gtrsim 10\ M_{\sun}\ \mathrm{yr^{-1}}$, 
e.g., \citealt{symeonidis11}; C13). 
For the first time, our results show that this 
linear relation remains applicable to galaxies
with SFRs down to 
$\sim 0.1\ M_{\sun}\ \mathrm{yr^{-1}}$\ 
in the redshift range of $0.5 \leq z < 2.0$\
(Sec.~\ref{sec:bhar_vs_sfr}).
The relation demonstrates that SMBH and galaxy growth
broadly track each other over cosmic time. 
This is consistent with the observational fact
that the evolutions of cosmic BHAR and SFR 
have broadly similar shapes. 
The normalization of cosmic SFR relative to 
cosmic BHAR is $\sim 3.7$~dex
(see, e.g., \citealt{silverman08, aird10, kormendy13}),
similar to our best-fit intercept of the 
\bharm-SFR relation.


The linear \bharm-SFR relation suggests a simple 
scenario of coevolution. H14 assumed that, 
for any individual galaxy, the ratio between the 
amount of gas accreted by its SMBH and that used 
to form stars is a universal constant
(when averaged over $\sim 100$~Myr). 
Under this assumption, the \bharm\ for different 
samples of similar SFR should be similar.
However, this assumption appears in contradiction 
with our observational results. 
For a given SFR level, the sources with larger $M_*$\
have significantly higher \bharm\ 
(Sec.~\ref{sec:bhar_vs_sfr_and_m}).
In addition, \bharm\ is also related to $M_*$\ 
linearly (Sec.~\ref{sec:bhar_vs_m}), 
and \bharm\ is correlated with $M_*$\ 
more strongly than SFR as indicated by our 
PCOR analyses in Sec.~\ref{sec:bhar_vs_sfr_and_m}.
Therefore, our results suggest that 
\bharm\ might be intrinsically linked to $M_*$, 
and this \bharm-$M_*$\ relation and the 
star-forming main sequence together might  
largely cause the observed \bharm-SFR relation
as a secondary effect. 

In the analyses of Sec.~\ref{sec:bhar_vs_sfr_and_m}, 
we find when controlling for $M_*$\ that the dependence 
of \bharm\ on SFR is relatively weak. 
We have furthermore checked the 
$\langle \mathrm{BHAR} \rangle$-sSFR relation for 
the whole sample, and do not find any significant trend.

\subsection{The Physical Link between BHAR and $M_*$}
\label{sec:bhar_m}
It has been well established that \xray-selected 
AGNs above a given \lx\ threshold 
are preferentially found in massive galaxies 
\citep[e.g.,][]{xue10, brandt15}. 
This finding is consistent 
with our results that \bharm\ depends 
strongly on $M_*$, even for SFR-controlled 
samples (Sec.~\ref{sec:bhar_vs_m} and 
\ref{sec:bhar_vs_sfr_and_m}). 
In fact, $M_*$, rather than SFR, appears to be 
the primary factor related to \bharm. 

Massive galaxies with $M_* \gtrsim 10^{10}\ M_{\sun}$\ 
have lower sSFR (Sec.~\ref{sec:bhar_ov_sfr_vs_m}) than 
less-massive galaxies. 
If we assume SFR reflects the amount of cold 
gas available, the decrease of sSFR for massive 
galaxies indicates the mass fraction of cold gas 
drops toward high $M_*$\ \citep[e.g.,][]{saintonge16}.
The cold gas needed for star formation is also likely 
responsible for fueling black-hole accretion
\citep[e.g.,][]{alexander12, vito14}, 
while hot-gas accretion could power 
low-luminosity AGNs that generally have little 
contribution to total black-hole growth
(see, e.g., 
Fig.~3 of \citealt{croton06,heckman14, yuan14}).
The $M_*$-dependent \bharm/\sfrm\ 
indicates the massive population is generally more 
efficient in feeding cold gas to their SMBHs 
(Fig.~\ref{fig:ratio_vs_M}).
This could be further broadly interpreted 
in two possible respects. 
First, the black-hole fueling efficiency of each 
galaxy might depend on $M_*$\ due to several 
physically plausible causes: 
\begin{enumerate}

\item The potential wells 
in galactic centers are deeper for massive galaxies,
making it easier for gas particles to fall into the 
galaxy center and fuel the SMBH.
More specifically, supernova feedback might prevent 
gas from falling into the galaxy center when the 
potential well is not sufficiently deep 
\citep[e.g.,][]{bellovary13, dubois15}.

\item Compared to low-$M_*$\ systems, 
high-$M_*$\ ones are more likely to have 
bars and major mergers 
\citep[e.g.,][]{melvin14, rodriguez-Gomez15}
that could induce gas inflow effectively 
\citep[e.g.,][]{alexander12}. 

\item Massive galaxies often have more 
massive SMBHs than low-$M_*$\ galaxies
\citep[e.g.,][]{kormendy13}.
Those SMBHs, having a stronger gravitational field, 
are more capable of accreting gas from their 
vicinity. In fact, some studies suggest 
long-term BHAR is proportional to \mbh, 
resulting from a universal Eddington-ratio 
distribution
\citep[e.g.,][]{aird12, jones16}.

\end{enumerate}
Second, the SMBH occupation fraction might drop toward 
low $M_*$\ ($M_* \lesssim 10^{10}\ M_{\sun}$), and 
some low-$M_*$\ galaxies might only host 
intermediate-mass black holes (IMBHs) with 
$M_{\rm BH} \lesssim 10^4\ M_\sun$\
\citep[e.g.,][]{volonteri10, miller15, trump15}.
Due to the Eddington limit, the \xray\ emission from 
accretion onto IMBHs is likely to be much weaker 
than that from SMBHs. The \bharm\ could thus 
be diminished for galaxies with lower $M_*$.

\subsection{Implications for the 
$M_{\rm BH}$-$M_*$\ Relation in the Local Universe}
\label{dis:mbh_m}
Our results have implications for the 
\mbh-$M_*$\ relation in the local universe,
and we illustrate this with some basic
arguments below.
The \mbh/$M_*$\ ratio for a galaxy at $z=0$\ can be 
estimated as
\begin{equation}\label{eq:mbh_ov_mstar}
\frac{M_{\rm BH} (t_0)}{M_* (t_0)} \approx
	\frac{M_{\rm BH} (t_2) + \int^{t_0}_{t_2} \mathrm{BHAR}(t) dt}
	     {M_* (t_2) + \int^{t_0}_{t_2} \mathrm{SFR}(t) dt},
\end{equation}
where $t$\ is cosmic time and the subscripts of $t$\ 
indicate redshift. Assuming that the mass of SMBH 
seeds is small compared to that accreted over 
cosmic history \citep[e.g.,][]{volonteri10} and 
most black-hole growth happens at 
$z \lesssim 2$,\footnote{This is broadly supported by 
the So{\l}tan argument, such that $\approx 70\%-80\%$\
of total black-hole accretion (including the contribution 
from luminous quasars) happens at $z \lesssim 2$\
\citep[e.g.,][]{ueda14, brandt15}. High-redshift 
luminous quasars mostly form SMBHs above $z\sim 2$\ 
\citep[e.g.,][]{wu15, banados16}, but they are rare objects 
and their discussion is beyond the scope of this study.}
we have 
$M_{\rm BH} (t_2) + \int^{t_0}_{t_2} \mathrm{BHAR}(t) dt 
\approx \int^{t_0}_{t_2} \mathrm{BHAR}(t) dt$, and
Eq.~\ref{eq:mbh_ov_mstar} can be simplified as 
\begin{equation}\label{eq:mbh_ov_mstar_A}
\frac{M_{\rm BH} (t_0)}{M_* (t_0)} \approx
	\frac{\int^{t_0}_{t_2} \mathrm{BHAR}(t) dt}
	     {M_* (t_2) + \int^{t_0}_{t_2} \mathrm{SFR}(t) dt}.
\end{equation}

For local giant ellipticals ($M_* \gtrsim 10^{11}\ M_\sun$), 
most of their stars are likely to have been formed at 
$z \gtrsim 2$\ \citep[e.g.,][]{chiosi02, siudek16}.
Thus, $M_*$\ is roughly the same over $z \approx 0-2$, i.e.,  
$M_* \equiv M_*(t_0) \approx M_*(t_2)$,
and Eq.~\ref{eq:mbh_ov_mstar_A} is 
approximately\footnote{For galaxy mergers, 
Eq.~\ref{eq:mbh_ov_mstar_ell} is still correct, 
providing mergers increase $M_{\rm BH}$\ and $M_*$\
proportionally.
}
\begin{equation}\label{eq:mbh_ov_mstar_ell}
\frac{M_{\rm BH} (t_0)}{M_*} \approx
	\frac{\int^{t_0}_{t_2} \mathrm{BHAR}(t) dt}
	     {M_*}\ \mathrm{(elliptical)}
	.
\end{equation}
If we assume the BHAR-$M_*$\ linear relation 
extends to $M_* \gtrsim 10^{11}$~$M_\sun$ and
has not evolved significantly
for \hbox{$z \lesssim 2$}\ (Eq.~\ref{eq:bhar_m} and 
Fig.~\ref{fig:Lx_vs_M}), 
Eq.~\ref{eq:mbh_ov_mstar_ell} leads to 
\begin{equation}\label{eq:mbh_ov_mstar_ellA}
\begin{split}
\frac{M_{\rm BH} (t_0)}{M_*} & \approx
	\frac{(t_2-t_0) \mathrm{BHAR}}
	     {M_*}\  \mathrm{(elliptical)} \\
	& \approx 10\ \mathrm{Gyr} \times 10^{-13}\
	       \mathrm{yr^{-1}} \\
	& \approx 10^{-3}. \\
\end{split}
\end{equation}
Therefore, the \mbh/$M_*$\ ratio for giant 
ellipticals should be approximately a constant
(1/1000), similar to the value (\hbox{$\approx 1/700$})
observed by \cite{haring04}.  
Considering that the existence of Compton-thick 
and broad-line AGNs could boost our observed 
\bharm\ (see Sec.~\ref{sec:missed_bhar}), 
the \mbh/$M_*$\ ratio might be several times 
larger and more consistent with the
value (\hbox{$\approx 1/300$}) from 
\citet[][but also see 
\citealt{shankar16}]{kormendy13}.

The above argument obviously depends on the 
assumption that giant ellipticals grow their
\mbh\ mostly at $z \lesssim 2$. 
This assumption, although under debate, 
is supported by observations of 
submillimeter galaxies (SMGs; see, e.g., 
Sec.~8.6.7 of \citealt{kormendy13}). 
SMGs are likely the high-redshift 
($z\sim2$) progenitors of massive 
ellipticals \citep[e.g.,][]{casey14, toft14}, 
and the growth of their SMBHs tends to lag that
of the host galaxies \citep[e.g.,][]{borys05, 
alexander08}. 
Nevertheless, it is also possible to reproduce 
the local $M_{\rm BH}/M_*$\ ratio 
if both the growth
of SMBHs and host galaxies take place at 
$z\gtrsim2$\ and have strong interplay. 
Simulations show that AGN feedback
can keep a tight 
$M_{\rm BH}$-$M_*$\ relation 
when both SMBHs and host galaxies grow at 
high redshift \citep[e.g.,][]{volonteri16}.

For local star-forming galaxies, 
most of their $M_*$\ is likely to be 
assembled at $z \lesssim 2$\ (e.g., B13). 
Thus, we have 
$M_*(t_2) + \int^{t_0}_{t_2} \mathrm{SFR}(t) dt 
\approx \int^{t_0}_{t_2} \mathrm{SFR}(t) dt$,
and Eq.~\ref{eq:mbh_ov_mstar_A} becomes
\begin{equation}\label{eq:mbh_ov_mstar_sf}
\begin{split}
\frac{M_{\rm BH} (t_0)}{M_* (t_0)} & \approx 
	\frac{\int^{t_0}_{t_2} \mathrm{BHAR}(t) dt}
	     {\int^{t_0}_{t_2} \mathrm{SFR}(t) dt} 
	\ (\varA{star-forming})\\
	& \sim \frac{\rm BHAR}{\rm SFR}  \\
	& \sim 10^{-5}-10^{-3.5} %
	,
\end{split}
\end{equation}
where we adopt the BHAR/SFR range from 
Figs.~\ref{fig:ratio_vs_M} and \ref{fig:ratio_vs_M_ms}.
Our results in Sec.~\ref{sec:bhar_ov_sfr_vs_m} show 
that long-term average BHAR/SFR positively depends
on $M_*$, and thus $M_{\rm BH} (t_0) / M_*(t_0)$\
is higher for massive galaxies than dwarf galaxies. 
This prediction is supported by some observations. 
\cite{miller15} and \cite{trump15} 
suggest that dwarf galaxies have 
a lower black-hole occupation fraction than 
massive galaxies, implying a generally lower 
\mbh/$M_*$\ for dwarfs. 
In addition, for some nearby galaxies
(e.g., M~33, NGC~205, and NGC~404), studies of 
nuclear kinematics place tight upper limits on 
\mbh, indicating the absence of SMBHs 
\citep[e.g,][]{gebhardt01, valluri05, nguyen16}.

Eqs.~\ref{eq:mbh_ov_mstar_ellA} and 
\ref{eq:mbh_ov_mstar_sf} suggest that ellipticals generally
have higher \mbh/$M_*$\ ratios than star-forming 
galaxies, consistent with some recent studies 
\citep[e.g.,][]{reines15, greene16}. 
From the viewpoint of this study, it is understandable 
why the observed local \mbh/$M_*$\ values 
($\sim 1/500$; e.g., \citealt{haring04, kormendy13}) 
are much higher than cosmic BHAR/SFR ($\sim 1/5000$; 
e.g., \citealt{silverman08, aird10}).
This is because local \mbh/$M_*$\ measurements are 
mainly based on observations of passive ellipticals
(Eq.~\ref{eq:mbh_ov_mstar_ellA}), and cosmic BHAR/SFR
is generally linked to \mbh/$M_*$\ of star-forming 
galaxies (Eq.~\ref{eq:mbh_ov_mstar_sf}).

From Eq.~\ref{eq:mbh_ov_mstar_ellA}, for massive ellipticals,
the \mbh/$M_*$\ is expected to be lower in the early 
universe (\hbox{$z \gtrsim 1$}); for star-forming 
galaxies, it should depend on $M_*$\ and have 
relatively weak cosmic evolution according to 
Eq.~\ref{eq:mbh_ov_mstar_sf}.
As discussed above, observations of SMGs 
support this scenario.
Some studies of high-redshift quasars find 
higher \mbh/$M_*$\ ratios than the local values, 
not expected in our scheme
(e.g., \citealt{ho07, merloni10};
but also see, e.g., \citealt{jahnke09, sun15}).
However, the \mbh/$M_*$\ measured from quasars might 
be biased and not representative for the majority 
of galaxies \citep[e.g.,][]{lauer07}. Also, large 
uncertainties often exist in the measurements of 
both \mbh\ and $M_*$\ for these quasars 
\citep[e.g.,][]{bongiorno12, shen13}.

\section{Summary and Future Prospects}
\label{sec:sum}
We have studied the dependence of SMBH growth on the 
SFR and $M_*$\ of host galaxies at $0.5 \leq z < 2.0$. 
Specifically, 
we compare \bharm\ for samples with different SFR 
and/or $M_*$. Due to the deep multiwavelength
data in the \cdfs, we are able to probe black-hole 
accretion in hosts down to $\mathrm{SFR} \sim 0.1\ 
M_{\sun}$~yr$^{-1}$ and $M_* \sim 10^{8}\ M_{\sun}$\
with reasonable completeness.
Our main results are summarized below:
\begin{enumerate}
\item \bharm\ correlates with SFR linearly 
      (Sec.~\ref{sec:bhar_vs_sfr}). However, for 
      SFR-controlled samples, galaxies with higher 
      $M_*$\ have higher \bharm\ 
      (Sec.~\ref{sec:bhar_vs_sfr_and_m}). 
      Thus, SFR does not appear to be uniquely 
      related to \bharm.
      The scenario in which long-term average BHAR 
      is only determined by host-galaxy SFR is 
      over simplified (Sec.~\ref{dis:coev}).  

\item \bharm\ is also proportional to $M_*$
      (Sec.~\ref{sec:bhar_vs_m}). 
      In fact, the correlation 
      between \bharm\ and $M_*$\ is stronger than that 
      between \bharm\ and SFR, suggesting $M_*$\ as  
      the primary host-galaxy property related to 
      SMBH growth. 
      This result also holds for the star-forming 
      population alone (App.~\ref{app:sf_galaxy}).
      The observed \bharm-SFR correlation 
      might be largely a secondary effect due to the 
      existence of the star-formation main 
      sequence (Sec.~\ref{dis:coev}). 

\item Massive galaxies 
      ($M_* \gtrsim 10^{10}\ M_{\sun}$)
      have higher \bharm/\sfrm\ ratios than their 
      less-massive counterparts 
      (Sec.~\ref{sec:bhar_ov_sfr_vs_m}), suggesting
      that they have higher black-hole fueling 
      efficiency and/or SMBH occupation fraction
      (Sec.~\ref{sec:bhar_m}).

\item Our results can naturally explain the observed 
      \mbh-$M_*$\ relation for local giant ellipticals, 
      and indicate that they have higher \mbh/$M_*$\ than 
      their progenitors in the earlier universe 
      (Sec.~\ref{dis:mbh_m}).
      Also, our results predict that \mbh/$M_*$\  
      for giant ellipticals is higher than that for
      star-forming galaxies in the nearby universe.
      Among local star-forming galaxies, 
      the \mbh/$M_*$\ values for massive galaxies 
      are likely to be higher than those for dwarfs.  
\end{enumerate}

In the future, this study could be extended to galaxies 
with larger $M_*$\ (SFR) by compiling a large number 
of luminous galaxies (Sec.~\ref{sec:bhar_vs_sfr}). 
To perform this investigation, analyses based 
on multiwavelength surveys of wider fields, 
e.g., COSMOS, XMM-LSS, and Stripe 82, are needed.
In addition, it is possible to extend this study 
to higher redshift using the \cdfs\ field, 
but this approach will require SED fitting that 
can eliminate potential high-redshift biases 
for SFR and $M_*$\ measurements (Sec.~\ref{sec:sfr}).
It would also be worthwhile to derive 
quantitative black-hole fueling efficiency and/or 
SMBH occupation-fraction estimates
as a function of $M_*$, based on the 
$M_*$-dependent \bharm/\sfrm\ 
(Sec.~\ref{sec:bhar_ov_sfr_vs_m}).
Future work could study the BHAR for giant
ellipticals by including morphological information
(Sec.~\ref{dis:mbh_m}) and the connection
between BHAR and host-galaxy gas content by using ALMA 
observations.

\acknowledgments
\section*{Acknowledgements}
We thank the referee for helpful feedback that improved 
this work. 
We thank Robin Ciardullo, Ryan Hickox, Luis Ho, 
Paola Santini, and Xue-Bing Wu for helpful discussions.
G.Y., C.-T.J.C., F.V., and W.N.B.\ acknowledge financial 
support from CXC grant GO4-15130A.
D.M.A.\ acknowledges the Science and Technology Facilities 
Council (STFC) for support through grant ST/L00075X/1.
B.L.\ acknowledges support from the National Natural Science 
Foundation of China grant 11673010 and the Ministry of 
Science and Technology of China grant 2016YFA0400702.
M.Y.S.\ and Y.Q.X.\ acknowledge support from the 
National Thousand Young Talents program, the 
973 Program (2015CB857004), NSFC-11473026, 
NSFC-11421303, the CAS Strategic Priority 
Research Program (XDB09000000), the Fundamental 
Research Funds for the Central Universities, 
and the CAS Frontier Science Key Research 
Program (QYZDJ-SSW-SLH006).
F.E.B.\ acknowledges support from 
CONICYT-Chile Basal-CATA PFB-06/2007 and 
FONDECYT Regular 1141218, and the Ministry of 
Economy, Development, and Tourism's Millennium Science 
Initiative through grant IC120009, awarded to The 
Millennium Institute of Astrophysics, MAS.
J.-X.W.\ acknowledges support from NSFC-11233002
and 973 program (2015CB857005).
This project uses Astropy (a Python package; see 
\citealt{astropy}).

\appendix
\section{Analyses for Star-Forming Galaxies}
\label{app:sf_galaxy}
Here we perform the same analyses as in Sec.~\ref{sec:res} 
but for star-forming galaxies only, in order to check if our 
conclusions also apply for this population alone.
We define star-forming galaxies as sources having 
sSFR within 5 times (i.e., 0.7~dex) of the median 
sSFR of the whole sample in the corresponding redshift 
range.\footnote{Different studies often adopt 
different empirical definitions of star-forming galaxies. 
Here we adopt a similar definition as \cite{elbaz11}, 
i.e., the galaxies with sSFR around a typical value.
We adopt a wider sSFR range than \cite{elbaz11} 
when defining the main sequence, mostly because our SED-based 
SFR estimations have larger uncertainties than the 
FIR-based SFR estimations in 
\citet[][see our Sec.~\ref{sec:sfr}]{elbaz11}.
} 
The median sSFRs for the low-$z$\ and high-$z$\ 
samples are $10^{-8.94}$\ and $10^{-8.72}$~yr$^{-1}$, 
respectively.
The numbers of star-forming galaxies in the low
and high redshift ranges are 8149 and 6604, 
respectively (see Tab.~\ref{tab:src_num_ms}).

\begin{figure*}[htb]
\includegraphics[width=0.52\linewidth]{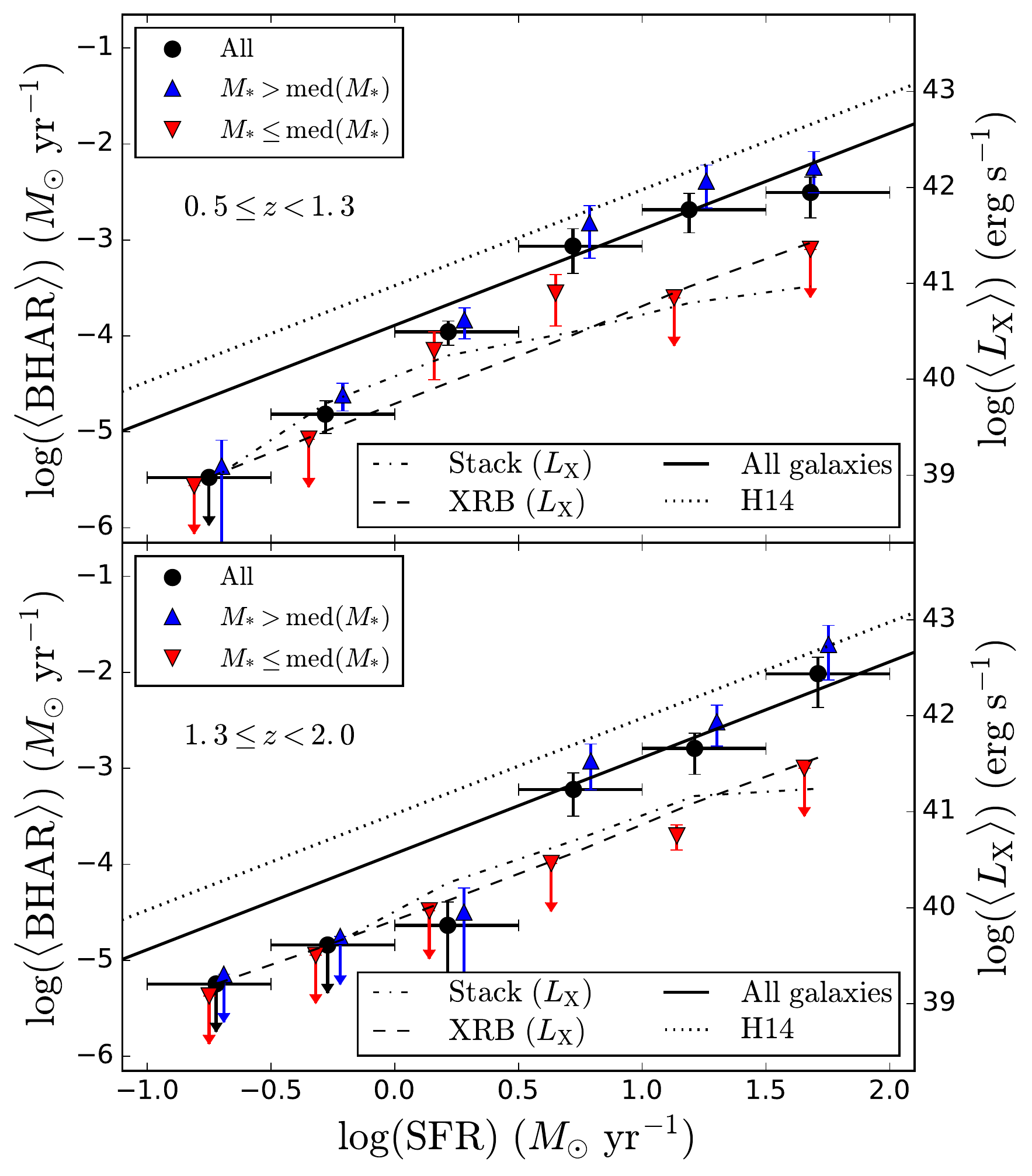}
\includegraphics[width=0.48\linewidth]{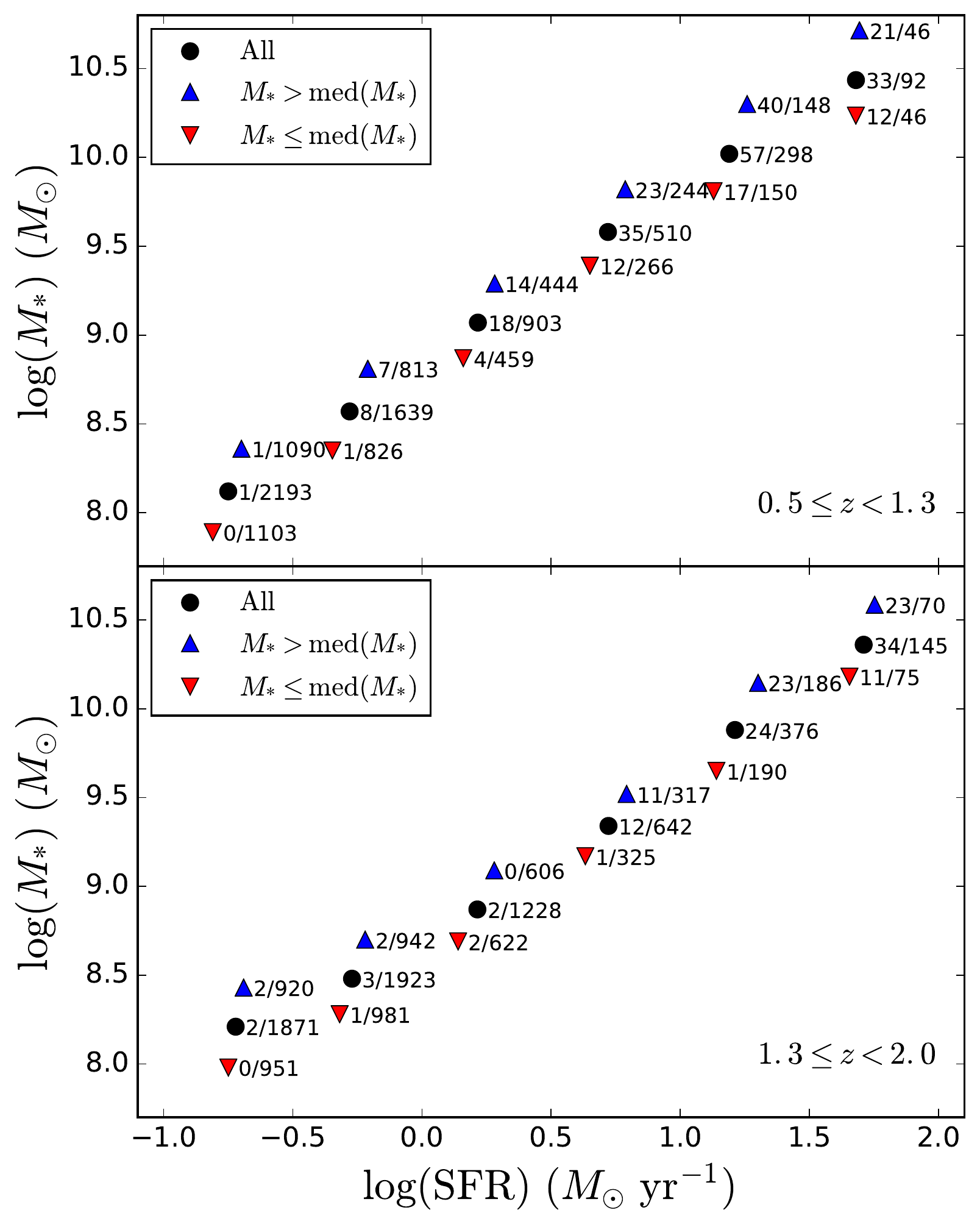}
\caption{Same format as Fig.~\ref{fig:Lx_vs_SFR} but 
for star-forming galaxies. The 
black solid line indicates the fitting of all galaxies as 
in Fig.~\ref{fig:Lx_vs_SFR}, rather than the fitting of 
star-forming galaxies.
}
\label{fig:Lx_vs_SFR_ms}
\end{figure*}

\begin{figure*}[htb]
\includegraphics[width=0.52\linewidth]{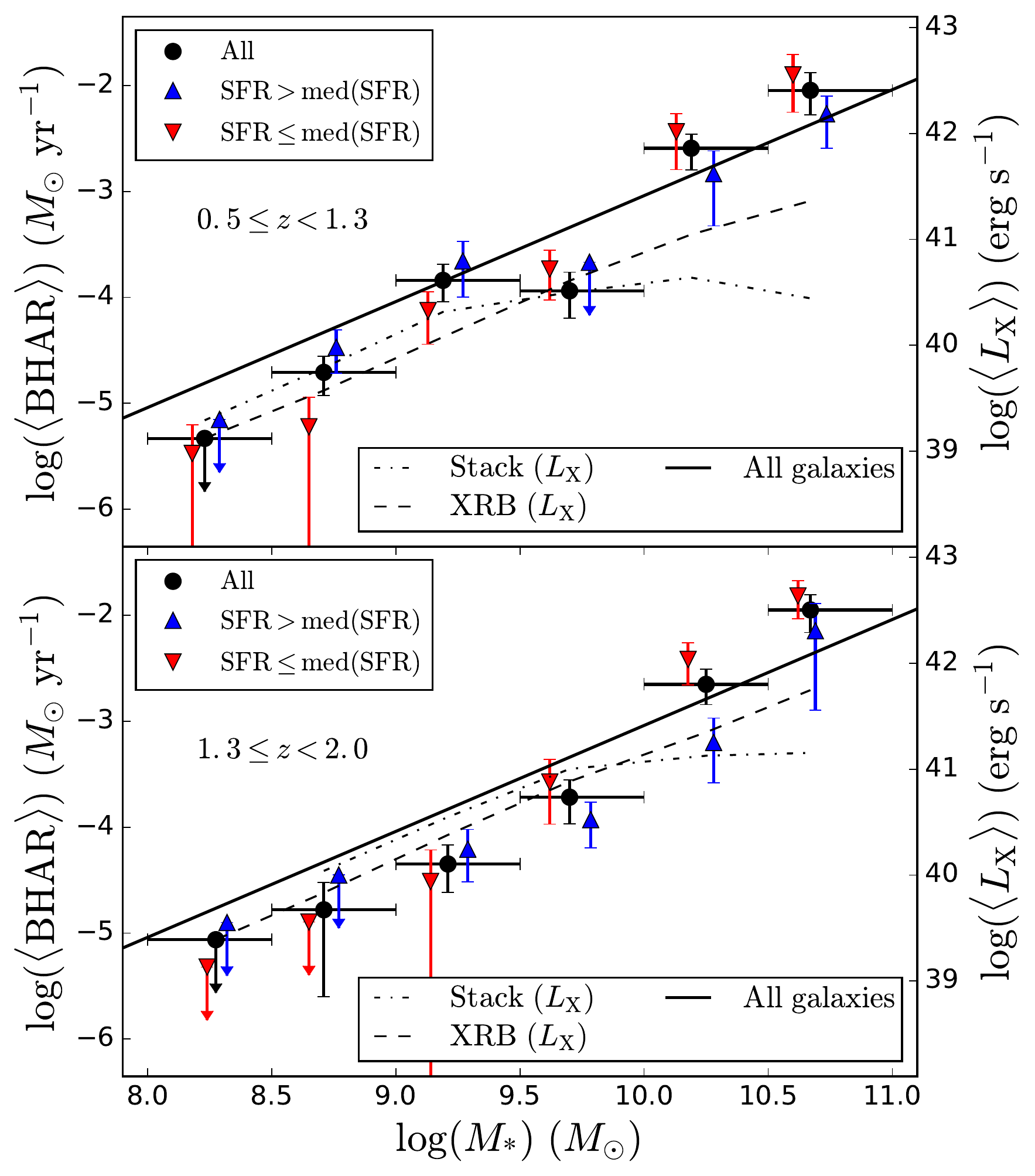}
\includegraphics[width=0.48\linewidth]{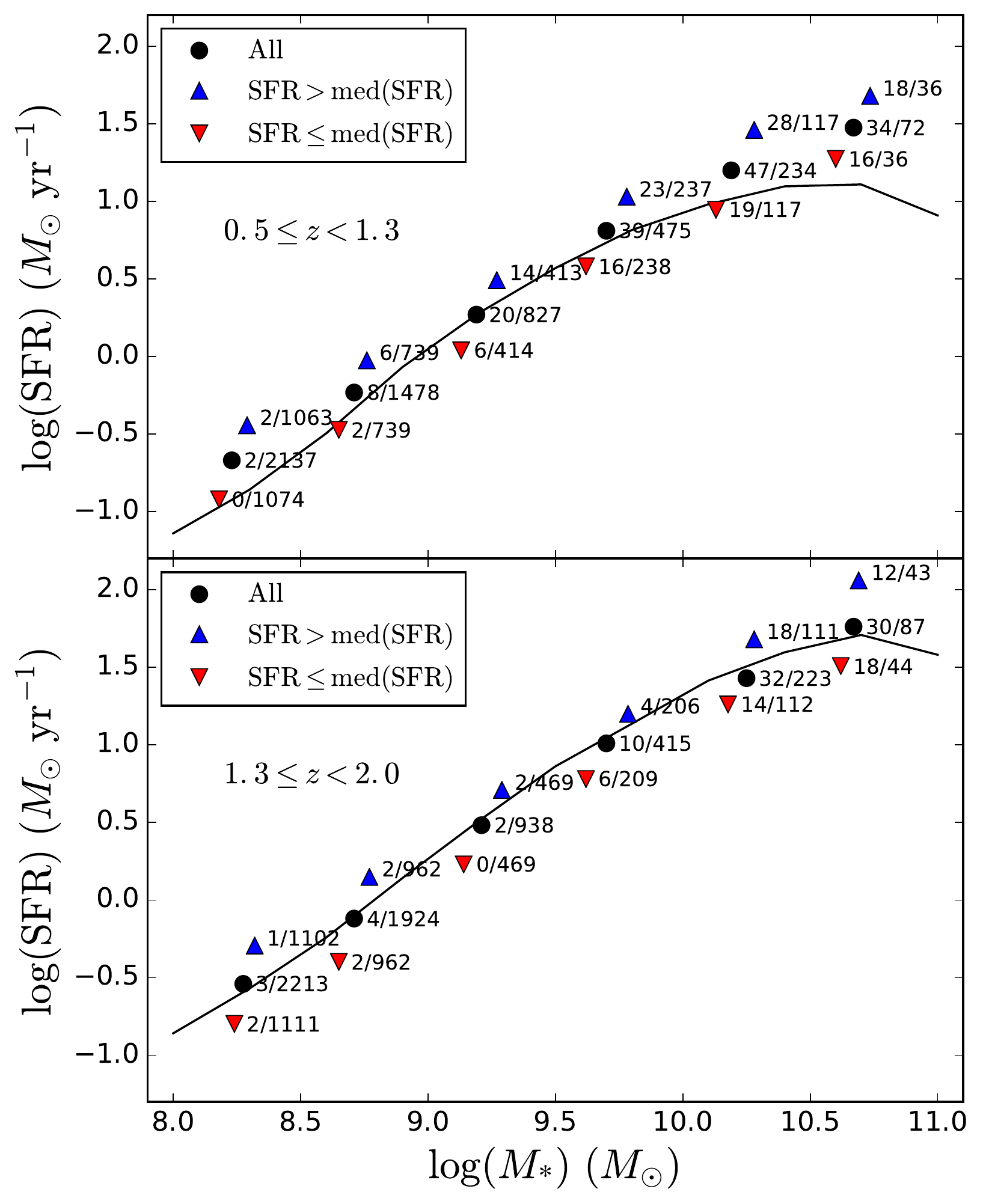}
\caption{Same format as Fig.~\ref{fig:Lx_vs_M} but 
for star-forming galaxies. The 
black solid line indicates the fitting of all galaxies as 
in Fig.~\ref{fig:Lx_vs_M}, rather than the fitting of 
star-forming galaxies.
}
\label{fig:Lx_vs_M_ms}
\end{figure*}

Similar to Sec.~\ref{sec:res}, we bin sources
based on SFR ($M_*$), and split each sample
based on $M_*$\ (SFR). For straightforward comparison with 
our results for all galaxies, we allow bins with
numbers of sources less than 100 in our analyses.
Figs.~\ref{fig:Lx_vs_SFR_ms} and \ref{fig:Lx_vs_M_ms} 
show the results. Similar to the results in 
Sec.~\ref{sec:bhar_vs_sfr} and \ref{sec:bhar_vs_m},
sources with higher SFR ($M_*$) generally have
higher \bharm. 
However, due to the apparent non-linearity, 
the linear models in Sec.~\ref{sec:bhar_vs_sfr} and 
\ref{sec:bhar_vs_m} result in unacceptable 
fitting quality, and thus we do not show the fitting 
results here.
From Fig.~\ref{fig:Lx_vs_SFR_ms} (Fig.~\ref{fig:Lx_vs_M_ms}), 
the non-linearity mainly arises from the apparent steep 
rise of \bharm\ above the threshold of 
$\mathrm{SFR}\sim 10^{0.5} M_\sun$~yr$^{-1}$\ 
($M_* \sim 10^{10} M_\sun$~yr$^{-1}$).
The steep change might be caused by statistical 
fluctuations atop a gradual rise, but it could also indicate 
an intrinsic threshold of SFR ($M_*$)\ above 
which AGN \xray\ emission becomes dominant over that of 
XRBs for star-forming galaxies. 
Larger samples are needed to differentiate the two 
possibilities.
The SFR-$M_*$\ relation for star-forming galaxies 
in Fig.~\ref{fig:Lx_vs_M_ms} 
(right) is more similar to that of B13 compared
to that in Fig.~\ref{fig:Lx_vs_M} at the 
high-$M_*$\ end (see Sec.~\ref{sec:bhar_vs_m}). 
 
From Fig.~\ref{fig:Lx_vs_SFR_ms}, the high-$M_*$\ 
subsamples generally have higher \bharm\ than the 
corresponding low-$M_*$\ subsamples.
From Fig.~\ref{fig:Lx_vs_M_ms}, the two subsamples 
with different SFRs generally have similar \bharm.
These results qualitatively agree with our major 
conclusion that \bharm\ correlates 
with $M_*$\ more strongly than SFR
(see Sec.~\ref{sec:bhar_vs_sfr_and_m}).
In Fig.~\ref{fig:Lx_vs_M_ms}, the low-SFR subsamples
of massive galaxies ($M_* \gtrsim 10^{10}$~$M_\sun$)
even appear to have higher \bharm\ than their high-SFR 
counterparts. However, the difference is not 
statistically significant, and larger
samples are needed to clarify this point. 
From PCOR analyses (see 
Sec.~\ref{sec:bhar_vs_sfr_and_m}), the parametric
method (Pearson) results in $\approx 2-3 \sigma$\ 
significances of the \bharm-$M_*$\ relation in the two 
redshift ranges in Tab.~\ref{tab:pca}, while 
the method shows the \bharm-SFR\ relation is 
insignificant in both redshift ranges. 
However, this parametric method models correlations  
linearly, which is likely not appropriate considering 
the apparent non-linearity in Figs.~\ref{fig:Lx_vs_SFR_ms}
and \ref{fig:Lx_vs_M_ms}. 
The non-parametric methods (Spearman and Kendall)
do not assume linearity but have less statistical
power (see Sec.~\ref{sec:bhar_vs_sfr_and_m}).
These methods cannot reveal high significances 
for either the \bharm-SFR or \bharm-$M_*$\ relations.
The reasons are likely to be the reduced sample 
size (especially the reduced number of AGNs; 
see Fig.~\ref{fig:m_vs_sfr}) and 
the reduced coverage of the SFR-$M_*$\ 
plane.

\begin{figure}[htb]
\includegraphics[width=\linewidth]{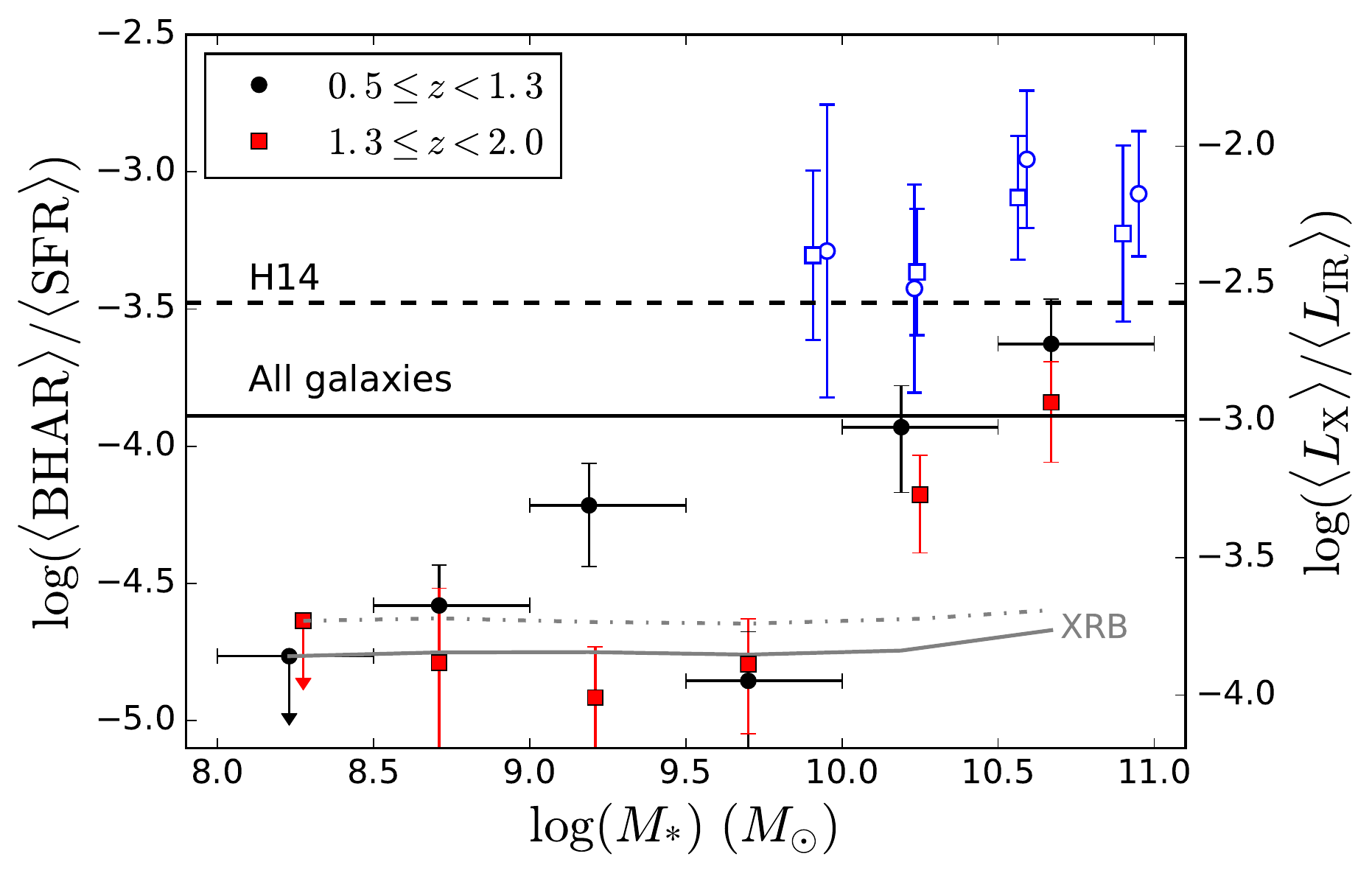}\\
\caption{Same format as Fig.~\ref{fig:ratio_vs_M} but 
for star-forming galaxies. The solid horizontal line 
indicates the best-fit intercept in 
Fig.~\ref{fig:Lx_vs_SFR}.\\
}
\label{fig:ratio_vs_M_ms}
\end{figure}

Fig.~\ref{fig:ratio_vs_M_ms} shows the ratio between
\bharm\ and \sfrm\ for the star-forming galaxies.
As in Fig.~\ref{fig:ratio_vs_M} for all 
galaxies, the ratio is higher for massive galaxies 
with $M_* \gtrsim 10^{10}\ M_\sun$. 
As expected from Fig.~\ref{fig:Lx_vs_M_ms}, 
there is a sudden ``jump'' of 
$\langle \mathrm{BHAR} \rangle$/$\langle \mathrm{SFR} \rangle$\ at 
$M_* \sim 10^{10}\ M_\sun$, probably 
indicating a physical $M_*$\ threshold 
above which AGN activity starts to become very strong.
Compared to the corresponding bins in 
Fig.~\ref{fig:ratio_vs_M}, \bharm/\sfrm\ values are 
slightly lower in general, although the trend
is weak and within the error bars. This is expected,
as the quiescent population has lower \sfrm\ 
but similar \bharm\ (our main conclusion) compared 
to the star-forming population at given $M_*$.

\begin{table*}
\begin{center}
\caption{Numbers of Sources in Different Samples of Star-Forming Galaxies}
\label{tab:src_num_ms}
\begin{tabular}{crrrc}\hline\hline
Sample & Low-$z$ & High-$z$ & Total & Fig(s). \\ 
(1) & (2) & (3) & (4) & (5) \\ \hline
All & 8149 & 6604 & 14753 &  N/A \\ 
$0.1 \leq \mathrm{SFR} < 100\ M_{\sun}$~yr$^{-1}$\ 
		& 5635 & 6185 & 11820 & 
		\ref{fig:Lx_vs_SFR_ms} \\ 
$10^{8} \leq M_* < 10^{11}\ M_{\sun}$\ 
		& 5223 & 5800 & 11023 & 
		\ref{fig:Lx_vs_M_ms} and \ref{fig:ratio_vs_M_ms} \\ 
\hline
\end{tabular}
\end{center}
{\sc Note.} --- Same format as Tab.~\ref{tab:src_num} but
only for galaxies near the star-forming main sequence.\\
\end{table*}

\bibliography{all.bib}
\bibliographystyle{apj}
\end{document}